\documentclass[12pt]{article}

\newcommand{\bfm}[1]{\mbox{\boldmath{$#1$}}}
\usepackage{graphicx}				
\usepackage{lineno}
\usepackage{xcolor}

\begin{document}

\title{
{An Energy-Angular Momentum Phase Function for Rubble Pile Asteroids\footnote{Accepted for publication in {\it Icarus}, 3/20/2025. \\ Author contact: scheeres@colorado.edu}}
}
\author{D.J. Scheeres \\ Smead Aerospace Engineering Sciences Department \\ University of Colorado Boulder}
\date{}
\maketitle

\begin{abstract}
This work analyzes the energetics of asteroid rubble piles in order to understand what asteroid morphologies should naturally arise from their formation and evolution process.
In doing this, a phase diagram is developed that maps out the range of final {minimum energy} states that a collapsing gravitational aggregate can achieve as a function of total angular momentum and mass distribution. This is developed assuming properties associated with rubble pile asteroids, and can provide insight into the formation and subsequent evolution of contact binaries and orbital binaries in the solar system as an outcome of catastrophic disruptions. 
The system angular momentum is used as an independent parameter, combined with resulting minimum energy configurations as a simple function of mass morphology of the final system. 
The configuration of systems with an energy boosted above the minimum energy state are also considered. 
This paper considers an ideal case, but outlines general results that can be continued for more precise models of distributed granular media modeled using continuum models or using discrete element models. 
\end{abstract}

\section{Introduction}

{The goal of this paper is to develop a simple model with supporting analysis that leverages the fundamental mechanics of rubble pile formation to outline the relationship between angular momentum, energy and binarity. A secondary goal is to provide specific tests for future numerical studies of the rubble pile asteroid formation process.} Resulting from this work is a phase function description of minimum energy states as a function of system angular momentum and mass distribution. This phase function is derived and analyzed and methods for its use are discussed, including as a way to track asteroid evolution.  
This is done by leveraging previous research performed by the author on the mechanics of rubble pile asteroids modeled as {collections of } rigid bodies \cite{scheeres_minE,scheeres_minE_chap,F3BP_scheeres}. Those earlier studies considered how specific situations and configurations would evolve as a function of system energy and angular momentum, but generally viewed each case in isolation or as a special case. In the current paper these earlier studies are used to develop a first principles approach to determining the likely and possible outcomes of a collapsing rubble pile solely as a function of its total angular momentum. The motivation for this work is to understand what the fundamental constraints are on the formation of rubble pile bodies and better identify the conditions that can lead to binarity in the resulting rubble pile asteroids. 

{This research is driven by the common occurance of binarity among small rubble pile asteroids. Recent estimates are that up to 30\% of small rubble pile asteroids are contact binaries \cite{virkki2022arecibo}, while up to 15\% of rubble pile asteroid systems are orbital binaries \cite{margot}. Thus almost half of rubble pile asteroids are predicted to have a bifurcated morphology of some sort. The ubiquity of this state of matter has only been emphasized by the recent Lucy flyby of Dinkinish \cite{levison2024contact}, which revealed that small body system to be a binary asteroid, and its secondary to itself be a contact binary. In this paper we explore the energetics of binary morphologies that could arise during the formation process of rubble pile asteroids. The current analysis is purely analytical, but should provide some theoretical insight into why binarity may occur so frequently in the solar system's population of small, rubble pile asteroids. }

The formation and evolution of rubble pile asteroid systems is generally modeled using numerically intensive discrete element models (DEM) that account for contact and gravitational forces between the elements \cite{sanchez_ApJ, schwartz_SSDEM, zhang2018rotational}. Such codes have their limitations as well, both in terms of the range of parameters which need to be specified, the time consuming nature of the computations, the limitations of the granular mechanics models, and the inherent numerical issues and computational errors which inevitably arise. Thus, the role of first principles models, such as developed in this paper, are important, as they can motivate a deeper understanding of the main principles at play, can yield analytical insights, and can motivate the numerical exploration of new regions of phase space. 

In this paper the evolution of a gravitational system that models the gravitational collapse that occurs during the formation of a rubble pile asteroid is analyzed. The model takes liberties in terms of the precise mechanisms of energy dissipation, strength and shape of accreted mass, and of timescales. However, some of the key assumptions made here for analytical tractability are frequently also made in numerical simulations -- such as the condensation of rubble piles into spherical collections of mass and the speed-up of dissipative effects \cite{michel2001collisions, michel2003disruption, nesvorny2010formation, nesvorny2021binary}. One key approach used in this analysis is allowing for the final, aggregated body to retain binarity in its mass distribution, motivated by previous studies studying the stability of single and binary collections of mass. Fundamental to this analysis is that the systems involved are micro-gravity aggregates, meaning that the strength of the rubble pile components are much greater than the gravitational pressures they feel \cite{holsapple_original, meyer_ApJL}. 

Previous research has investigated different analytical constraints for rubble pile asteroids as a function of angular momentum. {Holsapple studied the shape stability of given rubble pile bodies modeled using continuum mechanics theories. He was able to show that large ranges of asteroid ellipsoidal shapes could be stable just assuming the presence of internal friction and cohesion \cite{holsapple_original,holsapple_plastic,holsapple_spinlimits}.} Harris and Pravec \cite{pravec2007binary} modeled the total angular momentum of binary asteroids and related this to the disruption spin limit of all of the mass assembled into a single body. With this approach they show that many binary systems cannot be accreted into a single mass distribution without loss of excess angular momentum. Taylor and Margot \cite{taylor_binary_collapse} explored the minimum angular momentum that a binary asteroid requires in order to maintain its orbital state, and explored how this can affect a system's overall evolution. Scheeres \cite{scheeres_minE, scheeres_minE_chap, F3BP_scheeres} explored the mechanics of how multiple body systems evolve as their overall angular momentum evolves due to exogenous effects. All of these earlier papers are focused on explaining the current state of specific rubble pile systems and their evolution. In this paper the focus is on understanding the constraints on how binarity could naturally arise from the hypothesized formation mechanism of rubble pile asteroids. It is shown that a few key parameters will control the range of final states that can occur and that the precise final state that a system could settle into is not unique. 

This research is complementary to other work that has focused on the numerical simulation of planetesimal formation due to the streaming instability. In \cite{ nesvorny2010formation, nesvorny2021binary} a careful study is made of KBOs formed by gravitational collapse, with some similarities to the current work. Key results, which are also found here, are that the overall level of angular momentum fundamentally shapes the final, condensed systems that form. There are also key differences, beyond the fact that this paper is focused on analytical models. Also, as this study is focused on small rubble pile asteroids, the collapsing components are allowed to preserve elements of their shape, thus allowing for resting components for contact binary bodies. The analysis also tracks the final energy and angular momentum for the components modeled as rigid bodies. 

The paper is structured as follows. {In Section 2} the details of the model and the fundamental assumptions are outlined. {In Section 3} the different possible ends states for a collapsing system are reviewed and outlined as a function of angular momentum. In this section it is noted that there are two possible minimum energy final mass morphologies, that the matter collect into a single body or collect in a binary system with a given mass fraction. {In Section 4 a phase diagram is derived } expressing the minimum energy final morphologies as a function of angular momentum and mass distribution. {In Section 5 } the relative energies between possible final states are discussed, with a particular focus on the excited energy states of asteroids that have not settled into their minimum energy configuration. {In Section 6} we provide some example computations and map a number of existing asteroid systems into the phase diagram to highlight its possible use to track how exogenous effects can change the state of these systems. {Section 7 includes a discussion of the results and indicates future modifications to the theory that are of interest. Finally, Section 8 summarizes the conclusions from the study. }

\section{Model and Assumptions}

\begin{figure}[!ht]
    \centering
    \includegraphics[width=0.98\linewidth]{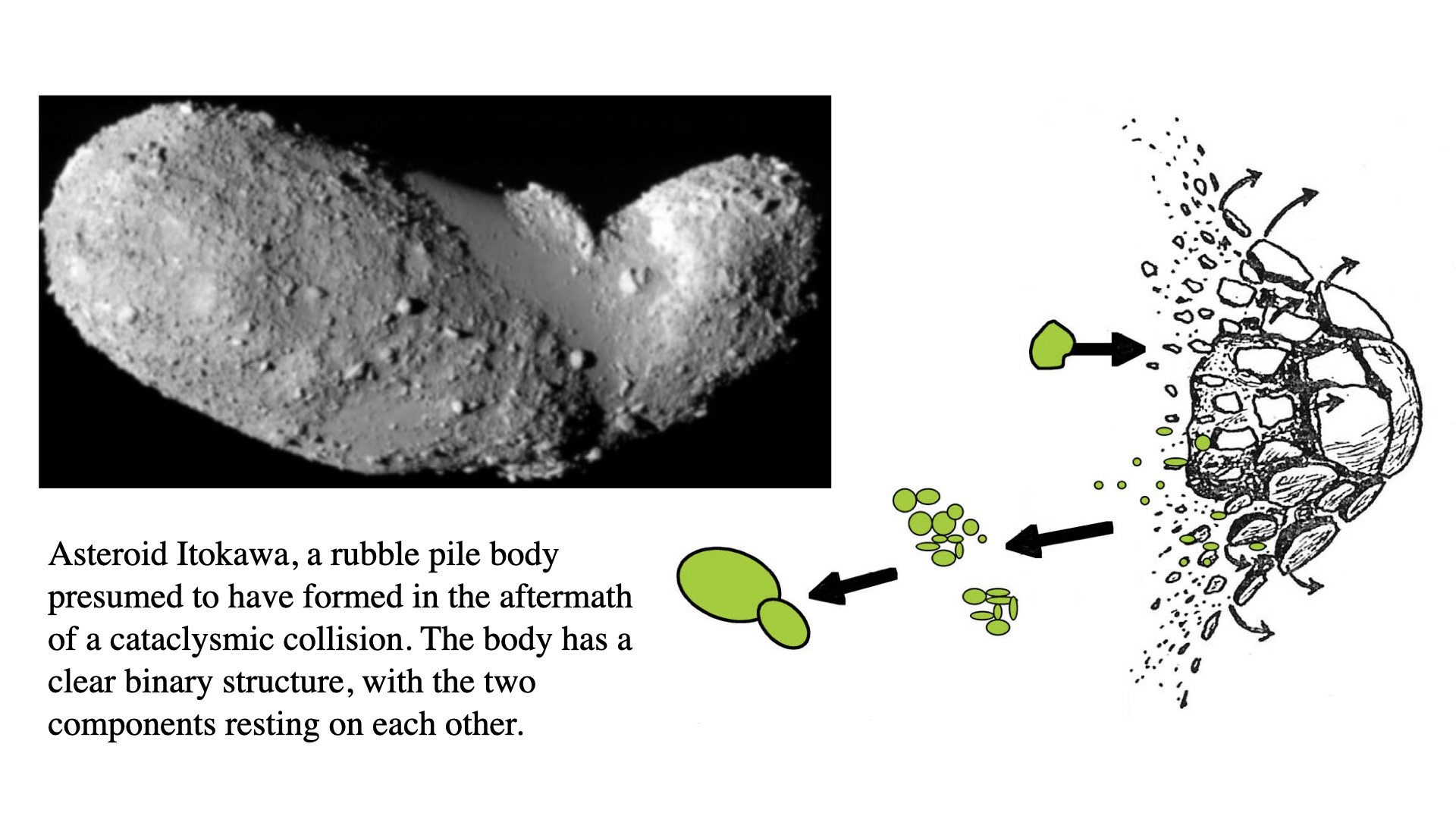}
    \caption{Presumed formation mechanism for the small rubble pile asteroid Itokawa \cite{fujiwara_science}. }
    \label{fig:formation}
\end{figure}

\begin{figure}[!ht]
    \centering
    \includegraphics[width=0.98\linewidth]{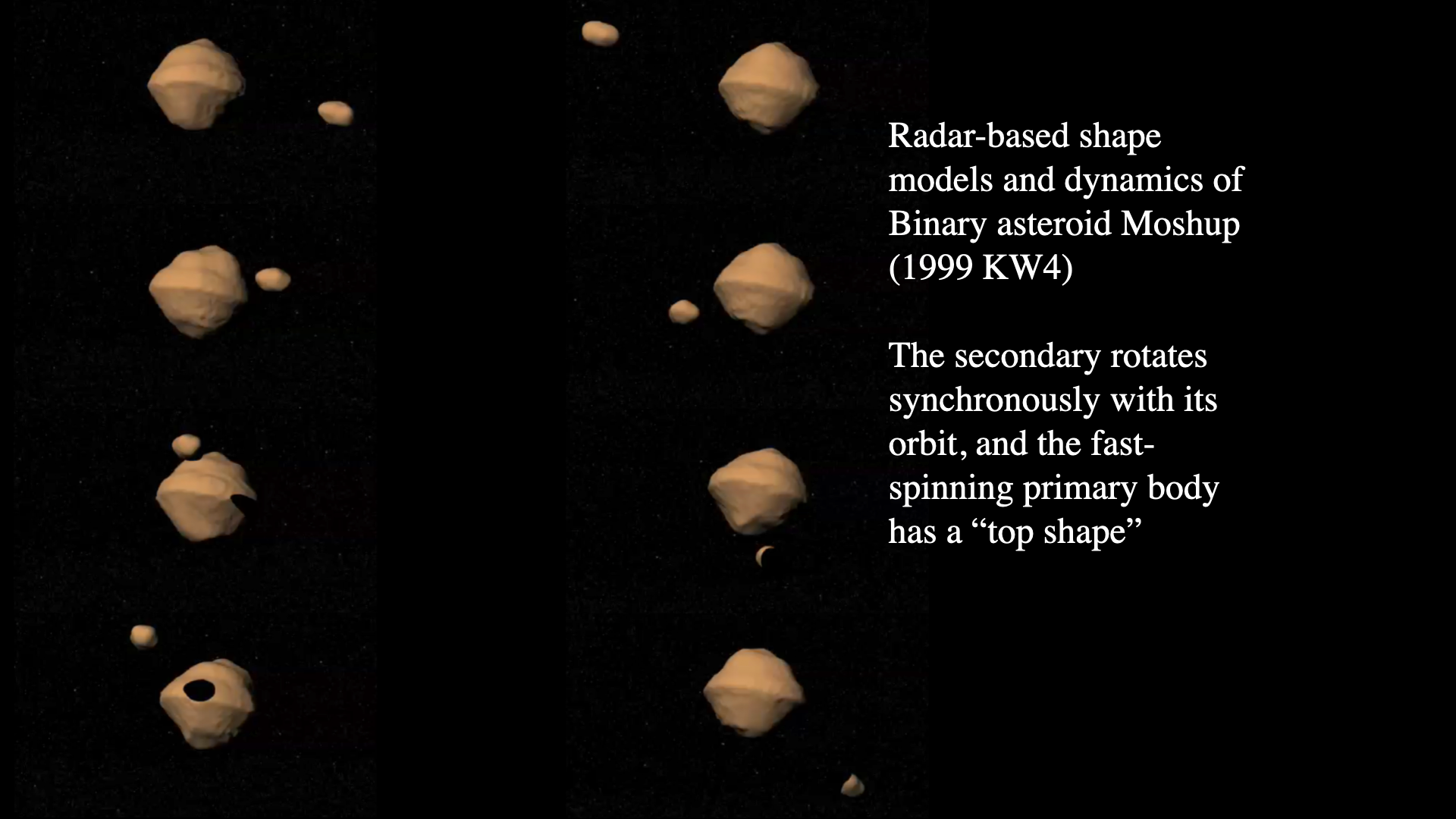}
    \caption{Radar-based model of binary asteroid Moshup \cite{KW4_ostro}. }
    \label{fig:1999KW4}
\end{figure}

This paper is motivated by the formation circumstances of small, rubble pile asteroids. In Fujiwara et al. \cite{fujiwara_science} the current state of the asteroid Itokawa is hypothesized to have arisen from the aftermath of a cataclysmic collision between two larger asteroids, as depicted in Fig.\ \ref{fig:formation}. While the formation of rubble pile bodies from such collisions has long been suspected \cite{michel2001collisions,michel2003disruption}, the Hayabusa mission to Itokawa provided the first tangible example of what such a system would look like, with the body clearly being a contact binary of two bodies, apparently formed separately before coming into contact. Similarly, in Fig.\ \ref{fig:1999KW4} are shown models of the binary asteroid Moshup (1999 KW4). Radar observations of this system provided the first clear picture of what a binary asteroid looks like, in terms of its primary and secondary system \cite{KW4_ostro}. Here the primary exhibited the top-shaped structure since seen in numerous fast-rotating asteroids with the secondary being in a synchronous state with its orbit. In the decades since these results, observations of other asteroid systems have only reconfirmed the basic morphology types shown in these rubble pile asteroids. 

\begin{figure}[!ht]
    \centering
    \includegraphics[width=0.98\linewidth]{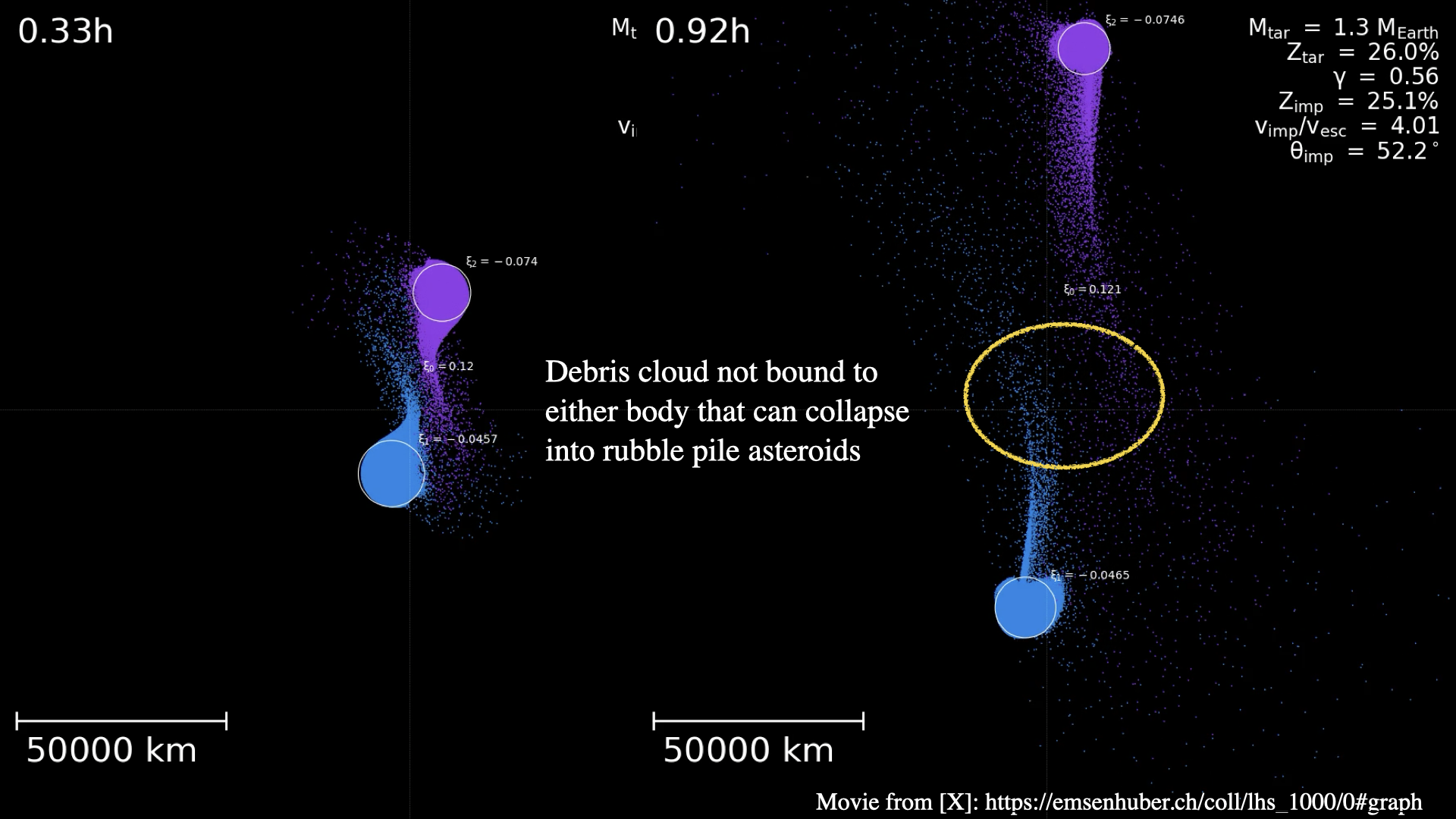}
    \caption{Snapshots of the aftermath of a glancing collision that could form rubble pile bodies \cite{emsenhuber2019fate}. }
    \label{fig:collision}
\end{figure}

Over the same time period, the modeling of catastrophic collisions has seen great advances due to both numerical method developments and insight into the range of different impact geometries and conditions that are most likely. These motivate the model developed here. A specific example can be seen in a simulation using Emsenhuber's website \cite{emsenhuber2019fate}, shown in Fig. \ref{fig:collision}. In this example, the aftermath of a hit and run collision, there will be streams of ejected material that are not re-accreted onto the larger bodies, but which depart the system in close proximity to each other and with similar speeds. For such a stream, conditions exist under which a portion of that stream can collapse and form a rubble pile body. The specific collapse process is complex and can involve ejection of bodies, collisions between bodies (at low speeds, however, and thus not catastrophic), and energy dissipation due to surface and impact forces, and thus requires detailed granular mechanics codes to model. However, under some realistic assumption we can find a range of strong constraints on how this entire re-accretion process plays out, and what the viable final morphologies of the system will be. 

\subsection{Specific Modeling Assumptions}

\begin{figure}[!ht]
    \centering
    \includegraphics[width=0.98\linewidth]{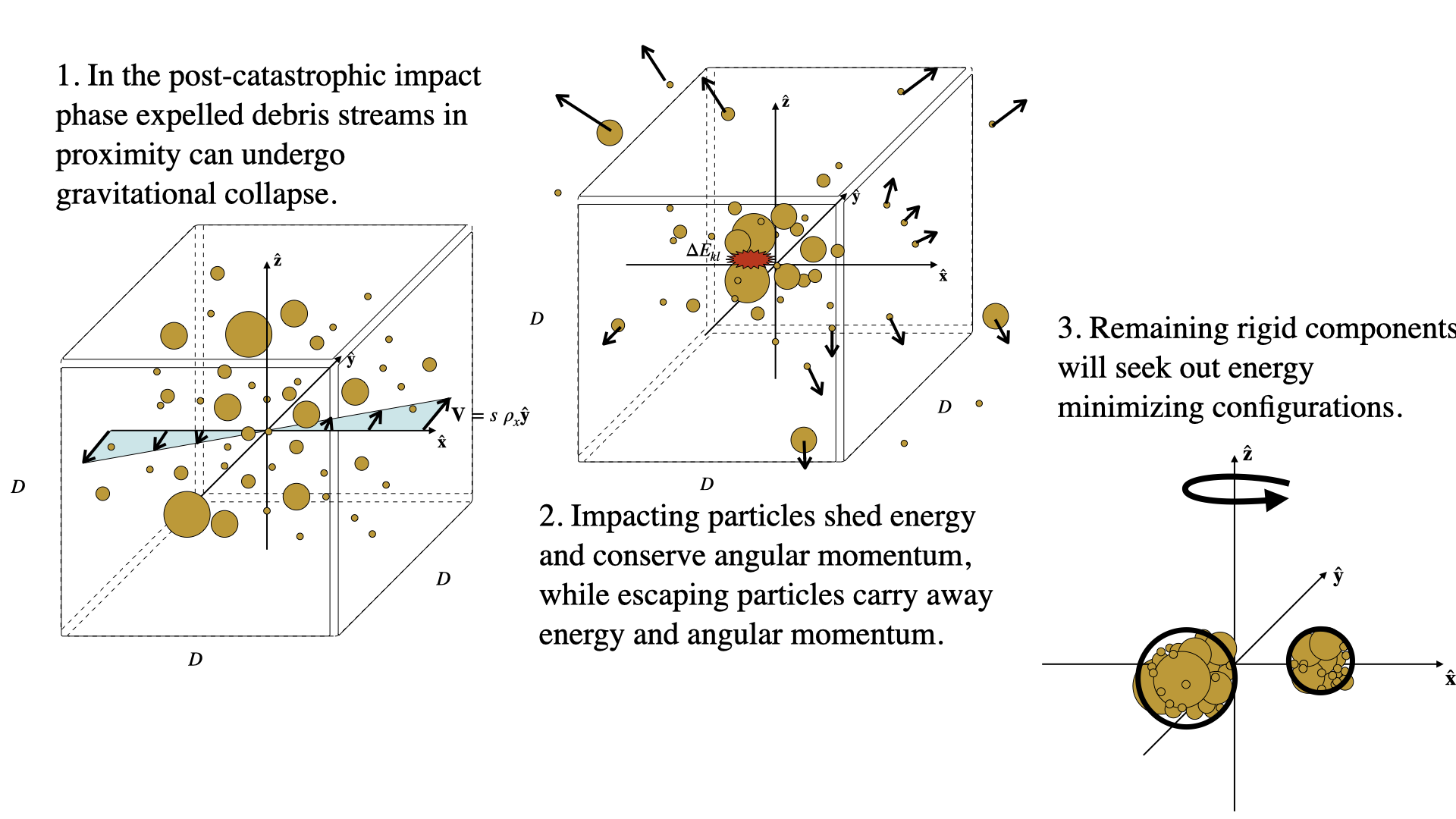}
    \caption{Cartoon showing the proposed formation sequence.  }
    \label{fig:formation_steps}
\end{figure}

The overall scenario assumed in our model is represented in Fig.\ \ref{fig:formation_steps}.  First, isolate a volume of distributed components that are gravitationally bound and which will gravitationally collapse into a stable, or semi-stable, state. 
If a collection of bodies is started in an initial distribution with relative velocities, the system will most likely lose bodies to gravitational ejection, which will tend to decrease the energy of the main distribution and sap angular momentum from the main system as well. An earlier study of this, for few body systems, shows that the remaining energy of the system will systematically decrease as bodies are ejected, but that the remaining angular momentum within clumps will be randomly distributed \cite{scheeres2023bounds}. 
This current study does not model this early phase of the evolution, but instead starts once there is a set of remaining particles that are in general bound and form a closed system, and can impact each other (at relatively slow speeds, meaning that one need not account for hypervelocity impacts). The relative impacts will dissipate energy and bring the system towards a final, stable state while conserving total angular momentum. In principle, one could imagine the final system achieving a minimum energy state, although as will be noted, there may be many other states at higher energies -- yet which are quasi-stable -- at which the system's evolution may stall. Once the impact phase of the collapse is over, if the system still has excess internal energy, the relative motion between components will continue to dissipate energy through tidal flexure of a single, rotating body and tidal forces acting between the components of a binary system. In this initial derivation, the effect of the sun is neglected. However, this is discussed later in the paper when some basic computations of the angular momentum and energy of a test distribution are computed. 

If the collection of masses evolve into a single body, then as a function of the interlocking of particles and parameters such as internal friction there can be infinities of different final shapes that the system can evolve into \cite{holsapple_original}. If the system were described as a continuum fluid without viscosity, then it would evolve towards a Maclaurin or Jacobi ellipsoid, however such an assumption for small rigid body asteroids has in general been seen to be inaccurate \cite{holsapple_original}. As the current paper strives to find analytic results that can shed qualitative light on this process, a strong assumption is introduced about the final collections of mass -- that they form a spherical distribution of matter. This allows us to simply account for the mutual and self gravitational potentials and to leverage previous analytical work on the stability of such finite density bodies when they come to rest relative to each other. To relax this assumption will likely require the use of numerical granular mechanics methods \cite{sanchez_icarus}, which is of interest for future analysis. Our current results will still assume that these spherical bodies (and their constituent particles) have an overall size and rotation and thus angular momentum and energy. 

\subsection{Analytical Model}

Consider an ideal situation with a volume $V$ containing a sparse size distribution of bodies, each with a mass $m_i$, an inertia tensor ${\bf I}_i$ and a finite density $\sigma_G$. The total mass of this system is then $M = \sum m_i$, and the total initial density is calculated as $M / V = \sigma \ll \sigma_G$. If each point in this distribution has a position and velocity relative to an inertial frame, we can formulate the system's internal energy and angular momentum by integrating over the system's relative positions and velocities. This was studied previously in \cite{scheeres_minE}, and yields
\begin{eqnarray}
	\bfm{H} & = & \frac{1}{2M} \int_{{\cal B}}\int_{{\cal B}} \left(\bfm{r} - \bfm{r}'\right)\times \left(\bfm{v} - \bfm{v}'\right) dm \ dm' \\
	E & = &  \frac{1}{4M} \int_{{\cal B}}\int_{{\cal B}} \left(\bfm{v} - \bfm{v}'\right)\cdot \left(\bfm{v} - \bfm{v}'\right) dm \ dm' + {\cal U} \\
	{\cal U} & = & - \frac{{\cal G}}{2}  \int_{{\cal B}}\int_{{\cal B}} \frac{dm \ dm'}{\left|\bfm{r} - \bfm{r}'\right|}
\end{eqnarray}
where ${\cal B}$ represents the complete particle distributions, {including the internal continuum distributions of each body with the differential mass element denoted by $dm$, and $\bfm{r}$ and $\bfm{v}$ are each element's position and velocity, respectively, with respect to the center of mass which is assumed to be at the origin. The three integrals presented are the total angular momentum vector of the distribution ${\bf H}$, the total energy of the distribution $E$, and the total gravitational potential energy of the distribution ${\cal U}$ (also including the self-potentials of the bodies). Note that the parameter ${\cal G}$ is the gravitational constant.}
Since the integrations also include the finite density regions of the bodies, this total angular momentum and energy also includes the rigid body rotational angular momentum and energy of each body as well \cite{scheeres_minE}. Further, the gravitational potential term also includes the self-potentials of each body. The total angular momentum of this closed system will then be conserved. The total energy will be reduced by the bodies impacting each other and dissipating excess energy through surface forces. If the bodies become collected into resting configurations in proximity to each other, then tidal forces can also lead to the dissipation of energy. While we do not model these dissipative forces, we implicitly assume their presence when we seek out the minimum energy configurations of such a system. 

Also important for our evaluations will be the total moment of inertia of the system relative to the angular momentum direction. This can be shown to equal
\begin{eqnarray}
	\bar{\bf I}_H & = &  \frac{1}{2M} \int_{{\cal B}}\int_{{\cal B}} \left[ \left(\bfm{r} - \bfm{r}'\right)\cdot \left(\bfm{r} - \bfm{r}'\right) - \left( \hat{\bfm{H}} \cdot \left(\bfm{r} - \bfm{r}'\right)\right)^2\right] dm \ dm'
\end{eqnarray}
{where $\bar{\bf I}_H$ is the complete inertia dyad for the system and} in this integral formulation includes the rigid body moments of inertia of each individual body. 
Here the vector $\hat{\bf H}$ is the unit vector along the angular momentum vector. This moment of inertia captures the total moment of inertia due to the relative placement of mass elements and also includes the rigid body inertias. This is not a constant, but is a general function of the relative positions and orientations of the mass distributions with respect to each other. 

A key function that ties together these three quantities of energy, angular momentum and moment of inertia is the amended potential \cite{smaleI, smaleII, scheeres2022derivation}, also referred to as the minimum energy function \cite{scheeres_minE}. This is a system-wide combination of angular momentum, potential energy and moment of inertia that is a strict lower bound on the system energy. We denote this function by ${\cal E}$, and it is defined as
\begin{eqnarray}
	{\cal E} & = & \frac{H^2}{2 I_H} + {\cal U} \\
	{\cal E} & \le & E
\end{eqnarray}
The lower bound expression is sharp, and the minimum energy function will equal the energy when the system is at a relative equilibrium. It can be shown that the difference $E - {\cal E}$ is the relative kinetic energy of motions that do not contribute to the total angular momentum, and thus when this is zero the system has no internal relative motions. This might occur at an instant only, but if it persists then the system must be in a relative equilibrium. In this paper most of our computations and constraints will involve this minimum energy function. 

We note a few important aspects of this function which are proven and explored in \cite{scheeres_minE}. First, if the angular momentum is fixed, then the expression is only a function of the relative configuration of the system and involves no velocity terms. Second, the gradients of this function with respect to its configuration variables are equal to zero at relative equilibria, providing a convenient way to compute and find these. Third, the energetic stability of a given relative equilibrium can be evaluated by inspecting the nature of the equilibrium states and the second order variations of the function evaluated at the equilibria. We note that energetic stability is a stronger condition than linear stability, as a system may be linearly stable but still able to dissipate energy. Such a relative equilibrium will not persist in nature, as it will eventually move away from its stationary state, even though the time for this migration can be long. Thus in this paper, when we call a system stable we mean that it is at its local minimum energy configuration. 

As noted above, the initial density of the system is $\sigma = M/V$ and must be less than the grain density of each body, $\sigma_G$. In the ensuing dynamical evolution we will be assuming that these bodies coallesce to form condensed bodies, with the grains at rest and in contact with their nearest neighbors. Such bodies will have an assumed bulk density greater than the initial density of material, but again less than the grain density, or $\sigma < \sigma_B < \sigma_G$. The recent missions to the rubble pile asteroids Itokawa, Bennu and Ryugu all determined bulk densities that were up to 1/2 of the expected grain densities, indicating a porosity approaching 50\%, or $\sigma_B \sim \frac{1}{2} \sigma_G$. Here we will assume the same, although our final results are not sensitive to this assumption. 

Finally, we can introduce non-dimensional quantities into our system, motivated by the above assumptions. Define the mass unit to be the total mass of the system, $M$. For the length unit, take the radius of the sphere consisting of all of the system mass with the defined bulk density $\sigma_B$. This is defined as $R = \left(\frac{3 M}{4\pi \sigma_B}\right)^{1/3}$. The time normalization is then computed as $\sqrt{R^3/{\cal G} M}$, and corresponds to the inverse orbital mean motion at the surface of this unit sphere. This is also the simple disaggregation spin rate of a spherical body (although note that when the geophysics are properly accounted for a body will start to fail at a lower spin rate \cite{hirabayashi2015failure}). Then the normalized angular momentum, energy and moment of inertia are defined as
\begin{eqnarray}
	\tilde{H}^2 & = & \frac{H^2}{{\cal G}M^3 R} \\
	\tilde{E} & = & \frac{E}{{\cal G} M^2 / R} \\
	\tilde{I}_H & = & \frac{I_H}{M R^2} 
\end{eqnarray}
with the non-dimensional minimum energy function equal to
\begin{eqnarray}
	\tilde{\cal E} & = & \frac{\tilde{H}^2}{2\tilde{I}_H} + \tilde{\cal U}
\end{eqnarray}


\section{Derivation of Minimum Energy End-States}

A fundamental mathematical result that we rely on is that the only absolutely stable end-states of a closed gravitational system are for it to condense into a single body, into two (or more) bodies resting on each other, or into two bodies {composed of aggregates} orbiting each other. This result can be derived from a theorem by Moeckel \cite{moeckel2017minimal} where it is proven that relative orbital equilibria of three or more rigid bodies in orbit about each other are always energetically unstable in the sense that they will never form a minimum energy configuration. Thus, even though triple and more system can form for rubble pile bodies, if the system dissipates sufficient energy it must always eventually reduce to a single or binary system. {This mathematical result can also be applied to rubble pile systems that can undergo internal shifts as they evolve. If the system has a fixed angular momentum, then any internal shifts will dissipate energy, bringing the system to a lower energy state. As discussed in \cite{scheeres_minE} such a system will eventually seek out a minimum energy configuration where where the entire system is in relative equilibrium, meaning that no relative motion occurs. At that point the Moeckel theorem will apply, meaning that a system with 3 or more bodies, even if starting in a relative equilibrium, will not be in a minimum energy state and is not fully stable in that there will always exist a pathway to a lower energy. }

Under our strong spherical final body assumption we will only consider three different end states, although we will comment on situations which deviate from this conservative definition later. 
\begin{enumerate}
\item
All of the mass is collected into a single spherical body
\item
All of the mass is collected into two spherical bodies resting on each other
\item
All of the mass is collected into two spherical bodies orbiting each other. 
\end{enumerate}
Under these restrictions there are rigorous results for which of these configurations can exist as a function of the total system angular momentum and as a function of how the mass is distributed between these two bodies. For the case of two spherical bodies resting on each other, we note a recent analysis \cite{meyer_ApJL} that shows that contact binary components only need a physically realistic cohesion or angle of friction in order to maintain their shape in such a contact binary configuration. 

Let us consider the relevant minimum energy stable states for each of these mass distributions in our non-dimensional units. We will split these discussions into the single and two body cases. 
At this point it is useful to remind the reader of the mutual gravitational potential between two spheres of mass $m_i$ and $m_j$ and distance $r_{ij}$, ${\cal U}_{ij}$, and the self-potential and moment of inertia of a sphere of mass $m_i$ and radius $R_i$, ${\cal U}_{ii}$ and $I_{H_i}$, respectively.  
\begin{eqnarray}
	{\cal U}_{ij} & = & - \frac{{\cal G} m_i m_j}{r_{ij}} \\
	{\cal U}_{ii} & = & - \frac{ 3 {\cal G} m_i^2}{5 R_i} \\
	I_{H_i} & = & \frac{2 m_i R_i^2}{5}
\end{eqnarray}
We also note that for rigid spheres we have the constraint that $r_{ij} \ge R_i + R_j$ with the equality occurring when the bodies are in contact. 
We derive non-dimensionalized results later when needed. 

\subsection{Single Body Case}
If all the mass is collected in a single body, then we have the following definitions for gravitational self-potential, moment of inertia and minimum energy function. Note that the mass of this sphere will equal the total mass and that the radius of this sphere will be, by definition, the unit length. These normalized quantities are then
\begin{eqnarray}
	\tilde{\cal U} & = & - \frac{3}{5} \\
	\tilde{I}_H & = & \frac{2}{5} 
\end{eqnarray}
and the amended potential is then
\begin{eqnarray}
	\tilde{\cal E}_{single} & = & \frac{5}{4} {\tilde{H}^2} - \frac{3}{5} \label{eq:singleE}
\end{eqnarray}

The single spherical body case is trivially in a relative equilibria. As the minimum energy function has no free variables, it is constant and equals the energy of the system. 
In this system the spin rate of the body can be found by taking the ratio of the angular momentum to the moment of inertia, or $\Omega = H / I_H$. Computing this for the non-dimensional form we find $\tilde{\Omega} = \frac{5}{2} \tilde{H}$. Dimensionalizing this expression gives us $\Omega = \sqrt{{\cal G}M/R^3}  \frac{5}{2} \tilde{H}$. For stability of the spherical body we require that the spin rate be less than the orbit rate at its surface, or $\Omega \le  \sqrt{{\cal G}M/R^3}$. Relating this to the angular momentum, we then find that the stability limit for a single spherical body can be stated as $\tilde{H} \le 2/5$, or, $\tilde{H}^2 \le 4/25 = 0.16$. 
In general, we will consider that if the normalized angular momentum is greater than the above limit, then the system cannot settle into a single body. If it is less than this limit, then this is one of the possible end states of the mass distribution. The disaggregation spin limit of $\tilde{\Omega} = 1$ will be used later when discussing excited energy states. 

\begin{figure}[!ht]
    \centering
    \includegraphics[width=0.98\linewidth]{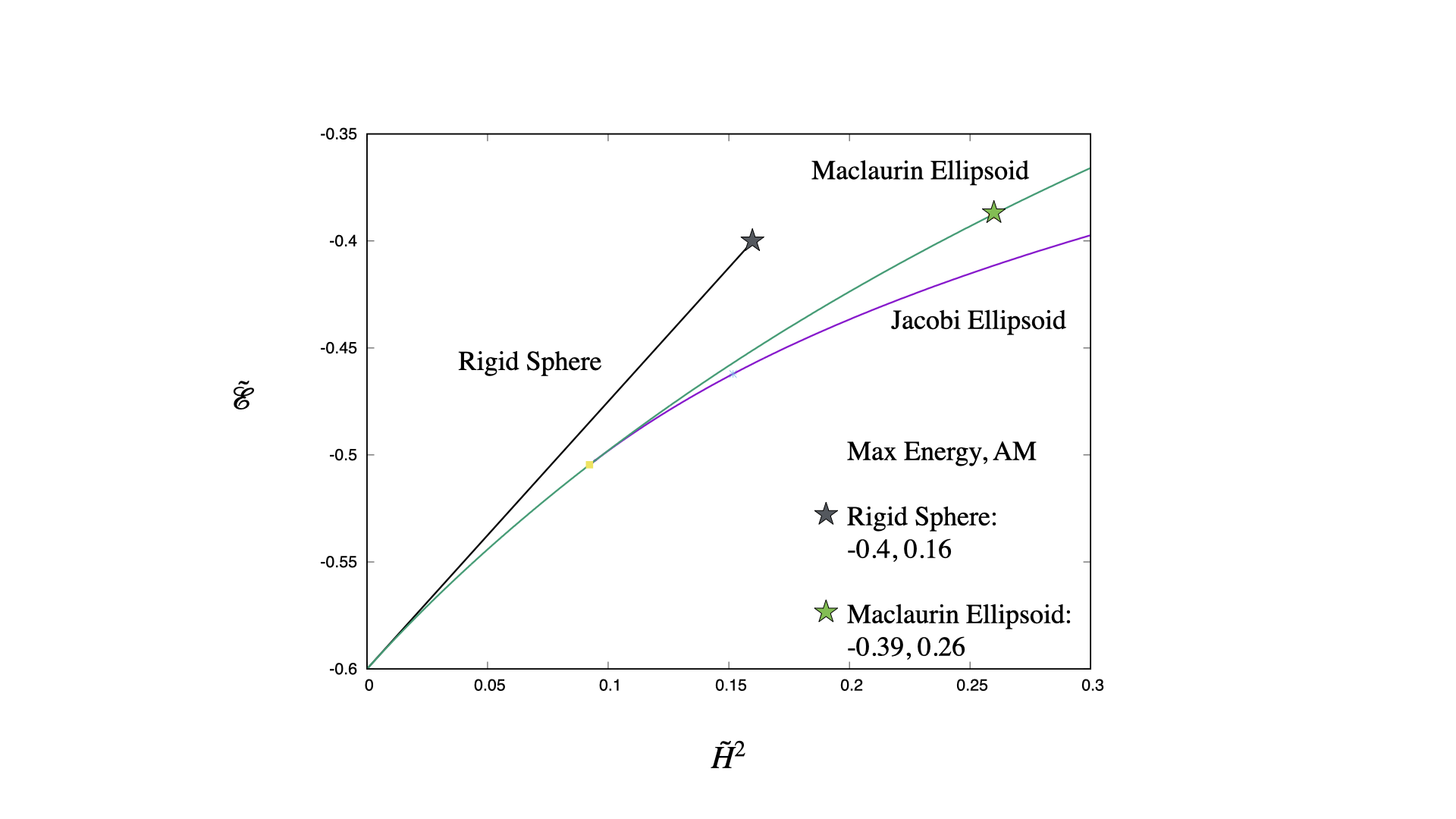}
    \caption{Chart showing the differences in energy and angular momentum between a spinning rigid sphere and the Maclaurin and Jacobi ellipsoids. {The stars signify the point where these bodies are at their maximum angular momentum and are no longer stable.}}
    \label{fig:maclaurin}
\end{figure}

It is important to note that if we allow for shape deviations from a spherical body, then in fact the stability limit will deviate from this result. As an example, in Fig.\ \ref{fig:maclaurin} we show the relative energy and angular momentum of a spinning sphere and of a classical Maclaurin and Jacobi ellipsoid, which are the true minimum energy shapes for a spinning distribution of mass. This provides an insight into the assumptions we are making with our current analysis, and indicates a direction for further improvement of the theory in the future. We note that a Maclaurin ellipsoid can carry about 0.1 units of additional angular momentum than a rigid sphere, however the total energy of the disaggregation spin state is nearly equal to the instability limit of the Maclaurin ellipsoid \cite{chandrasekhar1969ellipsoidal}, differing by only 0.01 units. 

\subsection{Two Body Case}
Next we assume that the mass is collected into two bodies, each of mass $m_i$ with radius $R_i = \left(\frac{3 m_i}{4\pi\sigma_B}\right)^{1/3}$. The normalized masses are $\mu_i = \frac{m_i}{M}$, and we have that $\mu_1 + \mu_2 = 1$. We then note that the non-dimensional radii have the simple definition $\tilde{R}_i = \mu_i^{1/3}$. We will take the convention that one mass is denoted as $0 \le \mu \le 1$ and the other mass is denoted as $1-\mu$. We denote the non-dimensional distance between the bodies as $\tilde{d}$ and note that it must be larger than or equal to the sum of the two spherical radii, denoted as $R_S(\mu)$. Thus, we have
\begin{eqnarray}
	\tilde{d} & \ge & R_S(\mu) \\ 
	R_S(\mu) & = & \mu^{1/3} + (1-\mu)^{1/3}
\end{eqnarray}

Then, the total potential and the total moment of inertia relative to the angular momentum are
\begin{eqnarray}
	\tilde{\cal U} & = & - \frac{\mu(1-\mu)}{\tilde{d}} - \frac{3}{5}\left[ \mu^{5/3} + (1-\mu)^{5/3} \right]\\
	\tilde{I}_{H} & = & \mu(1-\mu) \tilde{d}^2 \left[ 1 - (\hat{\bfm{d}}\cdot\hat{\bfm{H}})^2\right] +\frac{2}{5}\left[ \mu^{5/3} + (1-\mu)^{5/3} \right]
\end{eqnarray}
In the following we will assume that the two bodies lie in the plane orthogonal to the angular momentum vector, giving $\hat{\bfm{d}}\cdot\hat{\bfm{H}} = 0$. This will always be the minimum energy orientation of the two masses relative to the angular momentum, so we can make this assumption without loss of generality for the current study (later we relax this when considering non-minimum energy states). We also introduce the following function for the spherical moments of inertia for the system, $I_S(\mu)$
\begin{eqnarray}
	I_S(\mu) & = &  \frac{2}{5}\left[ \mu^{5/3} + (1-\mu)^{5/3}\right]
\end{eqnarray}
Note that the self potentials of the two spheres has a related definition of ${\cal U}_{S} = - \frac{3}{5}\left[ \mu^{5/3} + (1-\mu)^{5/3}\right] = - \frac{3}{2} I_S(\mu)$. 

Then the minimum energy function can be stated as
\begin{eqnarray}
	\tilde{\cal E} & = &\frac{1}{2}  \frac{\tilde{H}^2}{\mu(1-\mu) \tilde{d}^2 + I_S(\mu)}
		- \frac{\mu(1-\mu)}{\tilde{d}}  - \frac{3}{2} I_S(\mu)
\end{eqnarray}
where the distance between the two bodies $\tilde{d}$ is the only degree of freedom in this system. 

There are two situations to analyze, when the two bodies are in orbit about each other and when they rest on each other. 

\begin{figure}[!ht]
    \centering
    \includegraphics[width=1.1\linewidth]{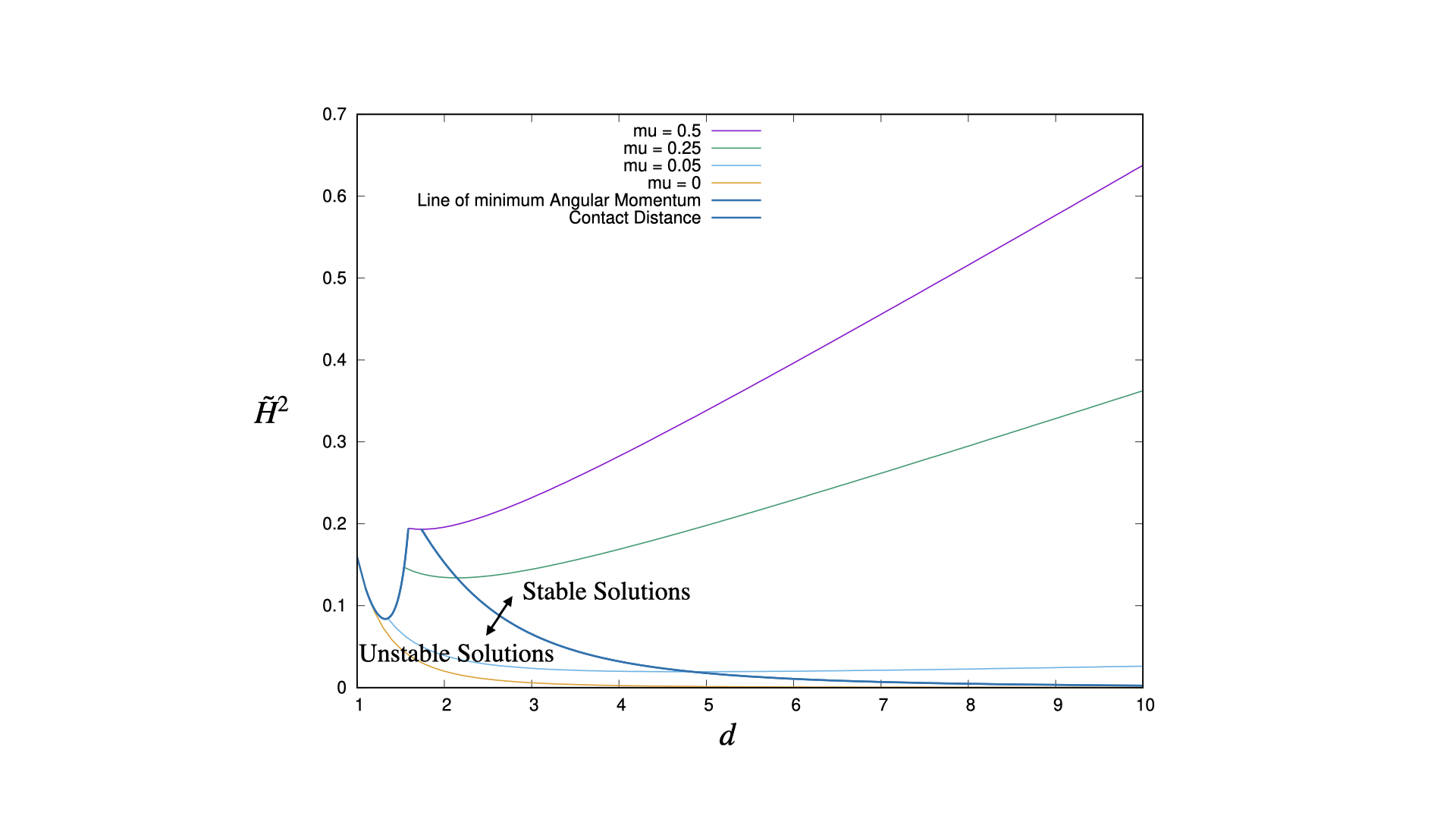}
    \caption{Angular momentum versus orbital distance, including regions of energetic stability and instability. {The line of angular momentum minima delineates the stable and unstable orbital equilibria.}}
    \label{fig:H2d}
\end{figure}

\paragraph{Orbiting Two Body Case}
To evaluate the orbital relative equilibria, we note that for this case the distance between the two bodies is a free degree of freedom. Taking the gradient of $\tilde{\cal E}$ with respect to $\tilde{d}$ gives us 
\begin{eqnarray}
	\frac{\partial\tilde{\cal E}}{\partial\tilde{d}} & = & 
		- \frac{\tilde{H}^2 \ \mu(1-\mu)\ \tilde{d}}{ \left[ \mu(1-\mu) \tilde{d}^2 + I_S(\mu)\right]^2} +  \frac{\mu(1-\mu)}{\tilde{d}^2} \label{eq:orbiteq}
\end{eqnarray}
and setting this equal to zero gives us the orbiting equilibrium condition. This can be solved for the angular momentum as a function of distance between the components \cite{scheeres_minE}
\begin{eqnarray}
	 \tilde{H}^2  & = & \frac{\left[ \mu(1-\mu) \tilde{d}^2 + I_S(\mu) \right]^2}{\tilde{d}^3}  \label{eq:H2eq}
\end{eqnarray}
We note that in this relative equilibria both spheres rotate at the same rate as the orbit, forming a doubly-synchronous system. 
Figure \ref{fig:H2d} shows the relation between $\tilde{H}^2$ and $\tilde{d}$ for a range of mass ratio values. Note the convex nature of this function, with there being a minimum in $\tilde{H}^2$ between the touching distance between the components, where the lines stop, and when the components are far from each other. This minimum angular momentum value defines the minimum angular momentum that a system with a given mass parameter can have. We will define this as the ``collapse'' angular momentum, a concept developed earlier \cite{taylor_binary_collapse}. We also note that there are orbital equilibria at all distances from touching to an arbitrary distance. Asymptotically we see that the angular momentum approaches $\tilde{H}^2 \sim [\mu(1-\mu)]^2 \tilde{d}$ for $\tilde{d}\gg 1 $ and $\mu \ne 0,1$. If we take the limit $\mu\rightarrow 0$ we find $\tilde{H}^2 = \frac{4}{25} \frac{1}{\tilde{d}^3}$, with a minimum distance of unity. Thus, we see that the stability condition for the single body matches with the orbit condition for an infinitesimal mass orbiting about the sphere. 

%
This minimum value of $\tilde{H}^2$ also plays a special role in the stability of the orbital solutions. To determine the stability of this equilibrium, we take the second partial of $\tilde{\cal E}$ with respect to $\tilde{d}$ and evaluate it at the equilibrium conditions. The details are worked out in \cite{scheeres_minE}, but we find that the energetic stability of the orbital solutions switches at the minimum of the angular momentum plot. The orbits are energetically unstable to the left of the minimum point in the curve and energetically stable to the right of them. This transition distance is 
\begin{eqnarray}
	\tilde{d}_C & = & \sqrt{\frac{3 I_S(\mu)}{\mu (1-\mu)}} \label{eq:d_C}
\end{eqnarray}
where the $C$ superscript denotes that this occurs at the collapse limit of the angular momentum. 
In Fig.\ \ref{fig:H2d} this distance is compared with the resting distance as a function of $\mu$, and we see that the orbits less than this distance are always energetically unstable. It is interesting to note that for the smallest companions, as $\mu \rightarrow 0$, that these orbits are asymptotically all energetically unstable. Thus, if a small particle is in orbit about a larger sphere and dissipates energy in the system, then its orbit will grow larger and larger. Also, if a system lies in a singly-synchronous orbit, with the primary spinning faster than the orbit rate,  the system will dissipate energy and grow to this stable relative equilibria. 

The minimum, or collapse, angular momentum for an orbiting case to exist for a given $\mu$ can also be solved for explicitly. 
This is found by evaluating the minimum value of angular momentum from Eqn.\ \ref{eq:H2eq}, which gives the minimum angular momentum for an orbit to exist at a given mass ratio, equal to
\begin{eqnarray}
	\tilde{H}^2_C & = & \frac{16}{3\sqrt{3}} \sqrt{I_S(\mu)}  \left[\mu(1-\mu)\right]^{3/2} \label{eq:H2Olim}
\end{eqnarray}
with the associated orbital distance given by $\tilde{d}_C$ in Eqn.\ \ref{eq:d_C}. 
This is plotted in Fig.\ \ref{fig:H2d} as the locus of the minima of the angular momentum curves. Each of the curves for a given value of $\mu$ intersect this curve at the point where orbits will bifurcate into existence under increasing $H^2$. If the angular momentum is less than this limit, the system cannot form an orbiting binary, and must either rest on each other or take on a different mass distribution as defined by $\mu$. 

This can be combined with the minimum distance at this angular momentum to define the collapse energy for an orbiting system as a function of $\mu$. If the system has an angular momentum less than this value, then the system cannot be in orbit and would collapse into a resting configuration. 

We can find an approximate relationship for the orbital distance as a function of angular momentum and mass fraction by performing an asymptotic expansion of Eqn.\ \ref{eq:H2eq}. A formal solution is
\begin{eqnarray}
	\tilde{d} & = & \frac{\tilde{H}^2}{\left[\mu(1-\mu)\right]^2} \left[ 1 - \sigma  - 3\sigma^2 - 15 \sigma^3 - 91\sigma^4 - 612\sigma^5 - 4389\sigma^6 - \ldots \right]^2 \\
	\sigma & = & \frac{I_s(\mu) \left[\mu(1-\mu)\right]^3}{\tilde{H}^4} 
\end{eqnarray}
with its derivation given in the appendix. 

With these results the minimum energy orbital solution can be stated as a function of angular momentum and mass ratio as:
\begin{eqnarray}
	\tilde{\cal E}_{orb} & = & \frac{\tilde{H}^2}{2\left[\mu(1-\mu)\tilde{d}^2(\tilde{H}^2,\mu) + I_S(\mu)\right]} - \frac{\mu(1-\mu)}{\tilde{d}(\tilde{H}^2,\mu)} - \frac{3}{2} I_S(\mu) \label{eq:Eorb}
\end{eqnarray}
under the conditions that $\tilde{H}^2 \ge \tilde{H}_C^2$ and $\tilde{d} \ge \tilde{d}_C$. 

\paragraph{Contact Two Body Case}
Next we evaluate the stability when the two bodies rest on each other. Here the mutual distance equals the sum of the radii, or $\tilde{d} =R_S(\mu)$. For this resting configuration to be stable, an infinitesimal increase in the distance between the two bodies must increase the energy. Thus, the loss of stability occurs when the change in energy with a change in distance is zero, which occurs when the orbiting condition reaches the contact distance. Evaluating the condition in Eqn.\ \ref{eq:H2eq} at the contact distance then gives us the limiting angular momentum at which the contact case becomes unstable, or when the system fissions, denoted by a subscript $F$.
\begin{eqnarray}
	 \tilde{H}_{F}^2  & = & \frac{\left[ \mu(1-\mu) R_S(\mu)^2 + I_S(\mu)\right]^2}{R_S(\mu)^3}  \label{eq:H2contactlim}
\end{eqnarray}
This is precisely the end of the orbital angular momentum curves in Fig.\ \ref{fig:H2d}, and indicates that when the resting configuration has an angular momentum equal to or greater than this value that it transitions into an orbiting system. For angular momentum levels less than this value, a resting configuration will always exist and is stable. 

This also provides an additional check on our stability condition for the single body, in terms of an infinitesimal mass on its surface. Taking the limit $\mu \rightarrow 0$ we find $\tilde{H}_F^2\rightarrow \frac{4}{25}$, in agreement with our previous limit. If the angular momentum is high enough then in general there will only be orbital cases possible. We can find a lower bound for this to be the case by evaluating the fission limit for $\mu = 0.5$, which we find to be equal to $\tilde{H}_F^2 = 2^{2/3} \ 49 / 400 \sim 0.1944\ldots$. For higher levels of angular momentum the only possible minimum energy configurations across all $\mu$ are orbiting solutions. 

The energy for this stable solution is then simply evaluated for an angular momentum $H^2 \le H_F^2$ as
\begin{eqnarray}
	{\cal E}_{rest} & = & \frac{H^2}{2 \left[\mu(1-\mu)R_S(\mu)^2 + I_S(\mu)\right]} - \frac{\mu(1-\mu)}{R_S(\mu)} - \frac{3}{2} I_S(\mu) \label{eq:Erest}
\end{eqnarray}

\section{Minimum Energy Phase Diagram}

With the above results, we can map out the possible final states of our system as a phase diagram relating energy, angular momentum and shape morphology. 
In addition, we can also evaluate systems which are excited away from the minimum energy configuration, if some assumptions are made on their excitation modes. The following subsections provide a detailed discussion of these results. 

Applying the above results, we find a phase diagram for the final morphology of the system, with the possible outcomes as a single body, a contact binary body, or an orbital binary, defined by the mass fraction between the two components, the minimum energy of these final morphologies and the system angular momentum. 
Using the system mass fraction $\mu$ to define its morphology, we can define a three dimensional graph of the minimum energy outcomes as a function of $\mu$, angular momentum $\tilde{H}$, and final energy $\tilde{\cal E}$. Below we present this phase space with a series of 2 and 3-dimensional plots. 

\subsection{Angular Momentum Transitions}

\begin{figure}[!ht]
    \centering
    \includegraphics[width=0.98\linewidth]{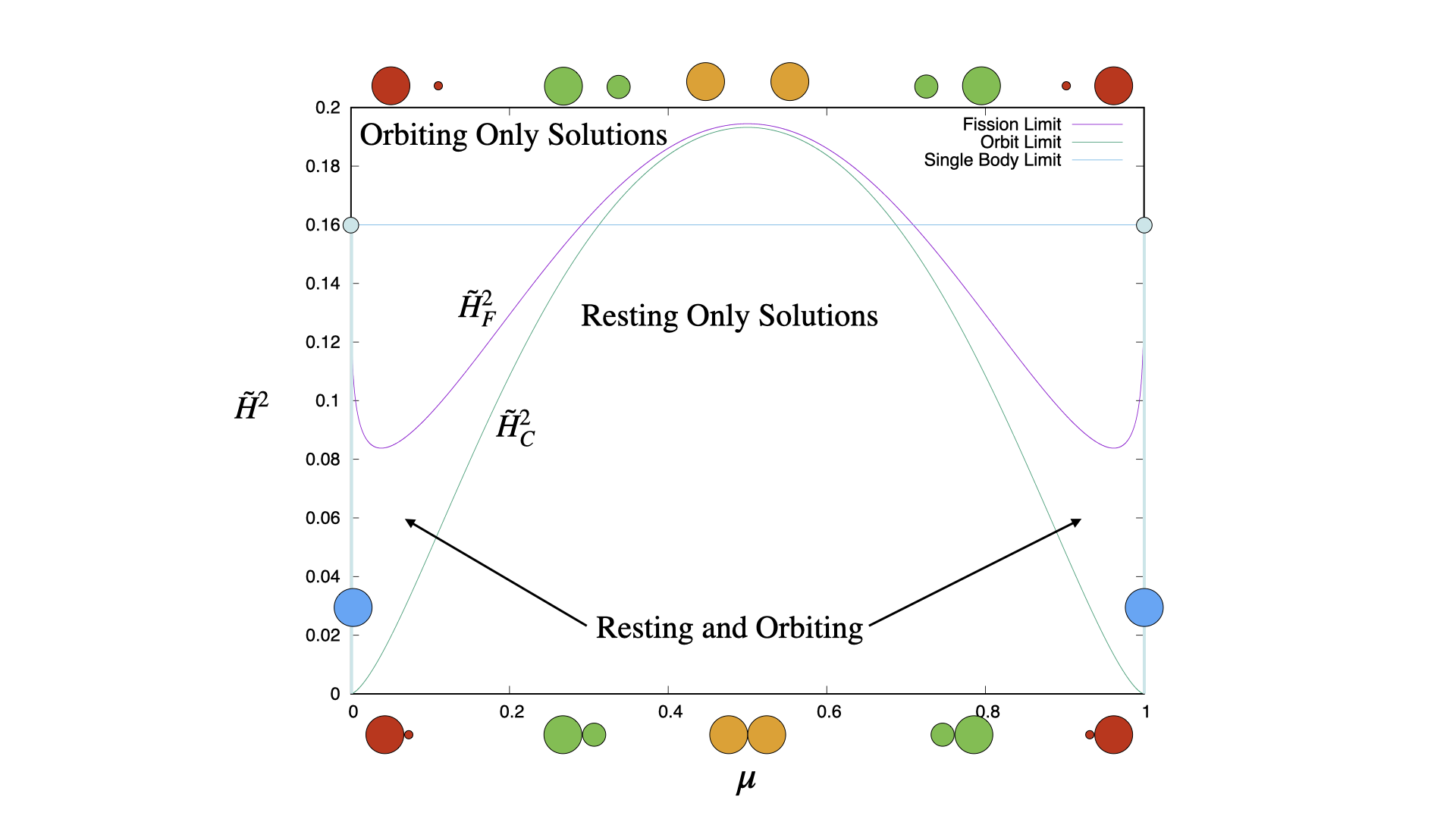}
    \caption{Phase diagram of stable end states as a function of angular momentum and mass fraction. }
    \label{fig:phase1}
\end{figure}

As a first step we evaluate the angular momentum limits as a function of mass ratio that lead to different minimum energy outcomes. The curves shown in Fig.\ \ref{fig:phase1} are the limiting angular momentum for orbits, $\tilde{H}_C^2$, and the limiting angular momentum for fission, $\tilde{H}_F^2$. Points with angular momentum below $\tilde{H}_C^2$ can only be resting, while points above $\tilde{H}_F^2$ can only be orbital. In between these limits both solution types exist. 

Figure \ref{fig:phase1} shows the phase diagram of different ``end state condensates'' as a function of angular momentum and mass fraction of the final binary system. This diagram shows the final conditions when all excess energy has been taken from the original system. 
Some particular points should be pointed out. For high enough angular momentum the initial closed system can only form an orbital binary as its minimum energy configuration. Conversely, for a low enough angular momentum for a given mass parameter we see that only contact binaries or singletons can exist. There are also regions where these states can all coexist, making the predicted final state of a given system ambiguous, appropriately capturing the stochastic nature of how rubble pile systems form. Of particular interest, we note that orbiting systems can exist for arbitrarily low angular momentum as $\mu\rightarrow 0$. We note that these systems will have the orbiting secondary at distances that become arbitrarily large as the mass ratio goes to zero. 

Not shown here are the final energies of these systems, which would show that at a given angular momentum, each of the final states will have an energy associated with it. If the system were allowed to arbitrarily shift mass around then it would  seek out the true minimum energy configuration for a given angular momentum. However, as we have pointed out earlier, at the low stresses present in rubble pile bodies, systems can easily rest on each other without undergoing significant reshaping, indicating that these can be stable configurations and would not necessarily be able to redistribute mass to seek out their absolute minimum energy distributions. Thus, in the realm of rubble pile asteroids this phase diagram can ideally be thought of as stable at every point. 

At this point we also note that our diagrams are symmetric in the shape parameter $\mu$. While we could truncate the plots to $\mu \le 0.5$, the full range of $\mu$ is kept to enable, in the future, the addition of non-spherical shapes for at least one of these bodies, which will cause these diagrams to be asymmetric. For an example of this see \cite{scheeres_F2BP_planar}. In light of this future work we keep the symmetry in these diagrams. 


\subsection{Energy Envelopes}

\begin{figure}[!ht]
    \centering
    \includegraphics[width=0.98\linewidth]{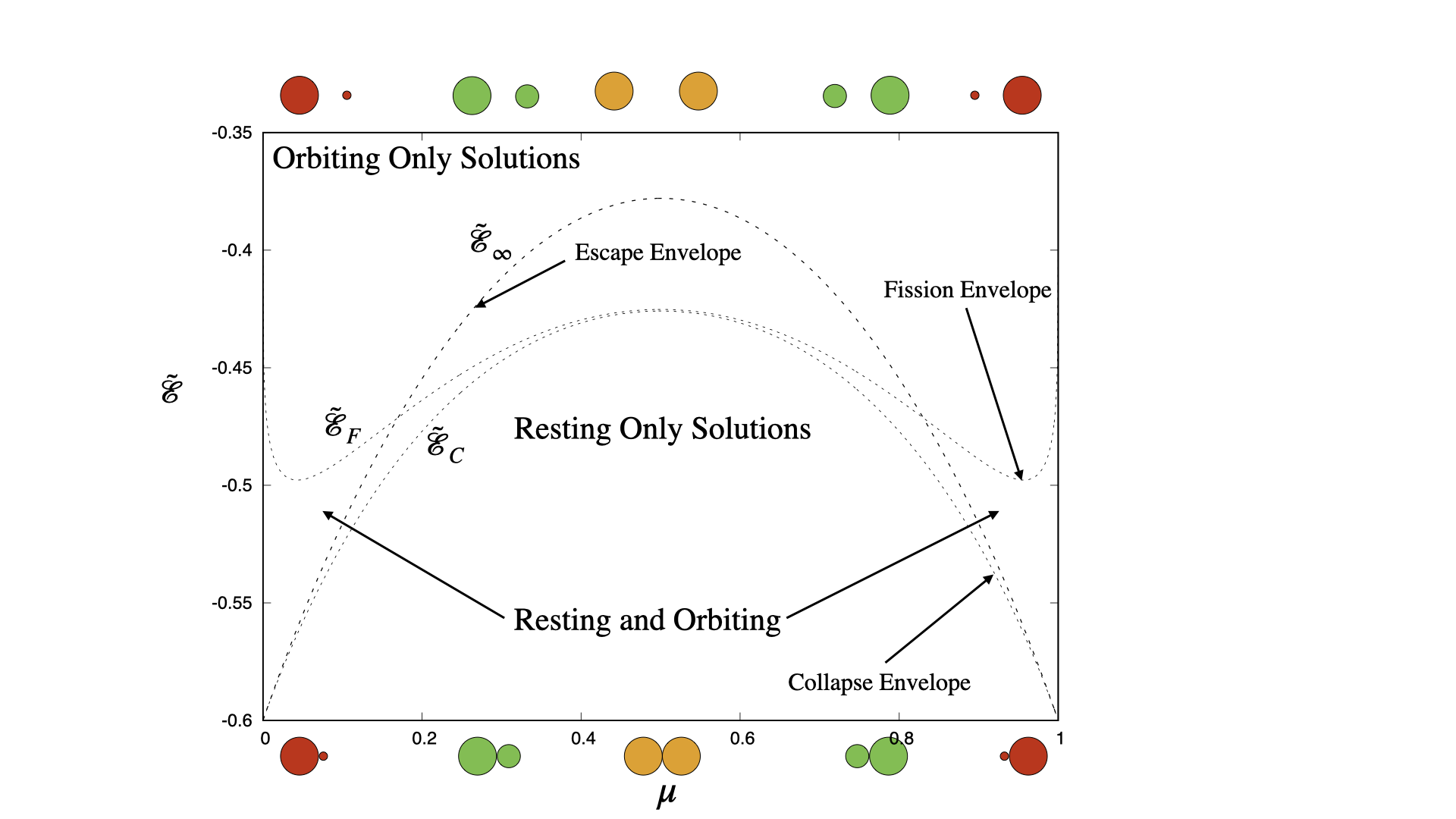}
    \caption{Chart of the energy envelopes for fission, collapse and escape. Above the fission envelope we can only have orbital configurations, within the collapse we can only have single or contact binary configurations, and outside of the escape envelop it is possible for binary systems to undergo escape. }
    \label{fig:envelope}
\end{figure}

Before looking at specific energy levels for given values of angular momentum, we plot the corresponding energy envelopes that exist between fission and collapse. As the angular momentum limits are only a function of mass ratio, when substituted into the corresponding energy formula the results are also only a function of mass ratio. Figure \ref{fig:envelope} plots the fission, collapse and escape envelopes, which are stated below. These define energetic limits as a function of mass fraction for different morphologies. 

The fission envelope is given by $\tilde{\cal E}_F$ and defines the energy, above which, a system must undergo fission. 
\begin{eqnarray}
	\tilde{\cal E}_F & = & - \frac{1}{2} \frac{\mu(1-\mu)}{R_S(\mu)} - \frac{1}{2} I_S(\mu) \left[ 3 - \frac{1}{R_S(\mu)^3} \right] 
\end{eqnarray}
Thus, all resting configurations can only exist below this energy limit. 

An important limit for the orbital case is found by evaluating this result at the minimum angular momentum and distance. This provides a minimum energy surface below which we cannot have two bodies in orbit. Evaluating this limit yields a lower energy bound as a function of $\mu$ only: 
\begin{eqnarray}
	\tilde{\cal E}_{C} & = & - \frac{\left[\mu(1-\mu)\right]^{3/2}}{3\sqrt{3 I_S(\mu)}} - \frac{3}{2} I_S(\mu)
\end{eqnarray}
This limit defines the collapse envelope, below which no orbital equilibria can exist. 

Finally, we also include the escape energy envelope, or the energy above which the system has sufficient energy for the two components to escape each other. This is simply computed as the orbit energy with the limit as $\tilde{d}\rightarrow\infty$, leaving the self potential as the remaining energy, or $\tilde{\cal E}_\infty = -\frac{3}{2} I_S(\mu)$. Any system with an energy above this satisfies the necessary conditions for escape, although not the sufficient conditions, meaning that it has sufficient energy to escape but may not do so -- depending on its current state. 

\subsection{Minimum Energy Phase Diagrams}

Next we consider the phase diagram of minimum energy states represented in the mass ratio -- energy space. 
We note that the energy can be a multi-valued function at a given value of angular momentum and mass ratio, meaning that multiple stable states of the system can exist at once. Then the relative energy of these define which is truly minimum energy and which is only a local minimum of the energy. We also note that each of these stable states have different ranges of angular momentum over which they are defined. 


For a single body, its minimum energy function is 
\begin{eqnarray}
	\tilde{\cal E} & = & \frac{5}{4}\tilde{H}^2 - \frac{3}{5}
\end{eqnarray}
for $\tilde{H}^2 \le \frac{4}{25}$, and is assumed not to be stable otherwise. 

For a binary body with a specified mass fraction $\mu$, the situation becomes more complicated. Here we borrow results from Theorem 7 in \cite{scheeres_minE}. First, when the contact equilibrium exists, the minimum energy function is given by Eq.\ \ref{eq:Erest} 
for $\tilde{H} \le \tilde{H}_F$. We note that $\tilde{\cal E}_{rest} \le \tilde{\cal E}_F$ with equality occurring when $\tilde{H} = \tilde{H}_F$. 


The orbital solution is already in existence when $\tilde{H} = \tilde{H}_F$, and can be traced out by parametrically plotting the angular momentum in Eqn.\ \ref{eq:H2eq} alongside the corresponding minimum energy function in Eq.\ \ref{eq:Eorb} which is evaluated at this angular momentum and distance. 

The orbital solution first comes into existence when $\tilde{H} \ge \tilde{H}_C$. Then it creates an inner, unstable solution that persists down to the fission limit given above. This unstable orbit is not shown in this figure, but is indicated in Fig. \ref{fig:2BP_EvsH2}. 
\begin{figure}[!ht]
    \centering
    \includegraphics[width=0.98\linewidth]{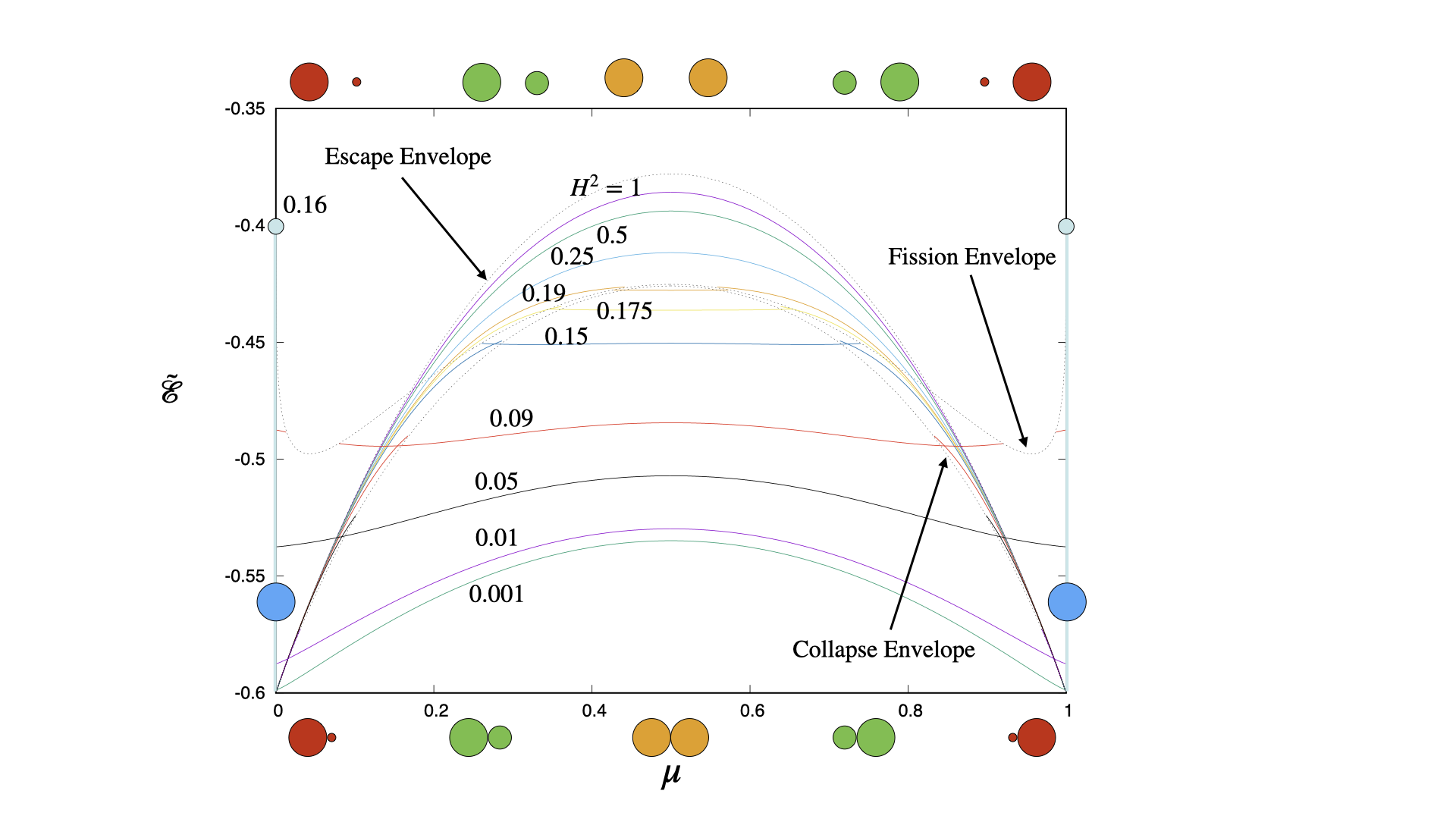}
    \caption{Phase diagram of stable end states as a function of energy and mass fraction, showing contours of constant angular momentum. }
    \label{fig:phase2}
\end{figure}

Figure \ref{fig:phase2} shows the phase diagram in terms of energy and mass fraction. To have a proper comparison with Fig.\ \ref{fig:phase1} we show contours of constant angular momentum on this diagram. This diagram is more complex, and requires more explanation. It still shows the ideal minimum energy configurations for a given value of angular momentum, but now there are three distinct families of curves to consider. The simplest are the vertical lines at $\mu = 0,1$ which represent the singleton case. Here the maximum energy equals $-0.4$ and corresponds to the ideal disaggregation limit. Next, the more horizontal lines that appear for angular momentum values less than $\sim0.2$ correspond to resting equilibria. Finally, the inverted-parabolic curves at higher values of angular momentum correspond to orbital binary minimum energy states. At high angular momentum the orbital curves are defined across all mass ratios, while at low angular momentum the resting curves are defined across all mass ratios. At moderate levels of angular momentum these two types of curves are no longer defined across all mass ratios, and at a given angular momentum we see that there can be multiple values of energy for the different final configuration morphologies. 

\begin{figure}[!ht]
    \centering
    \includegraphics[width=0.98\linewidth]{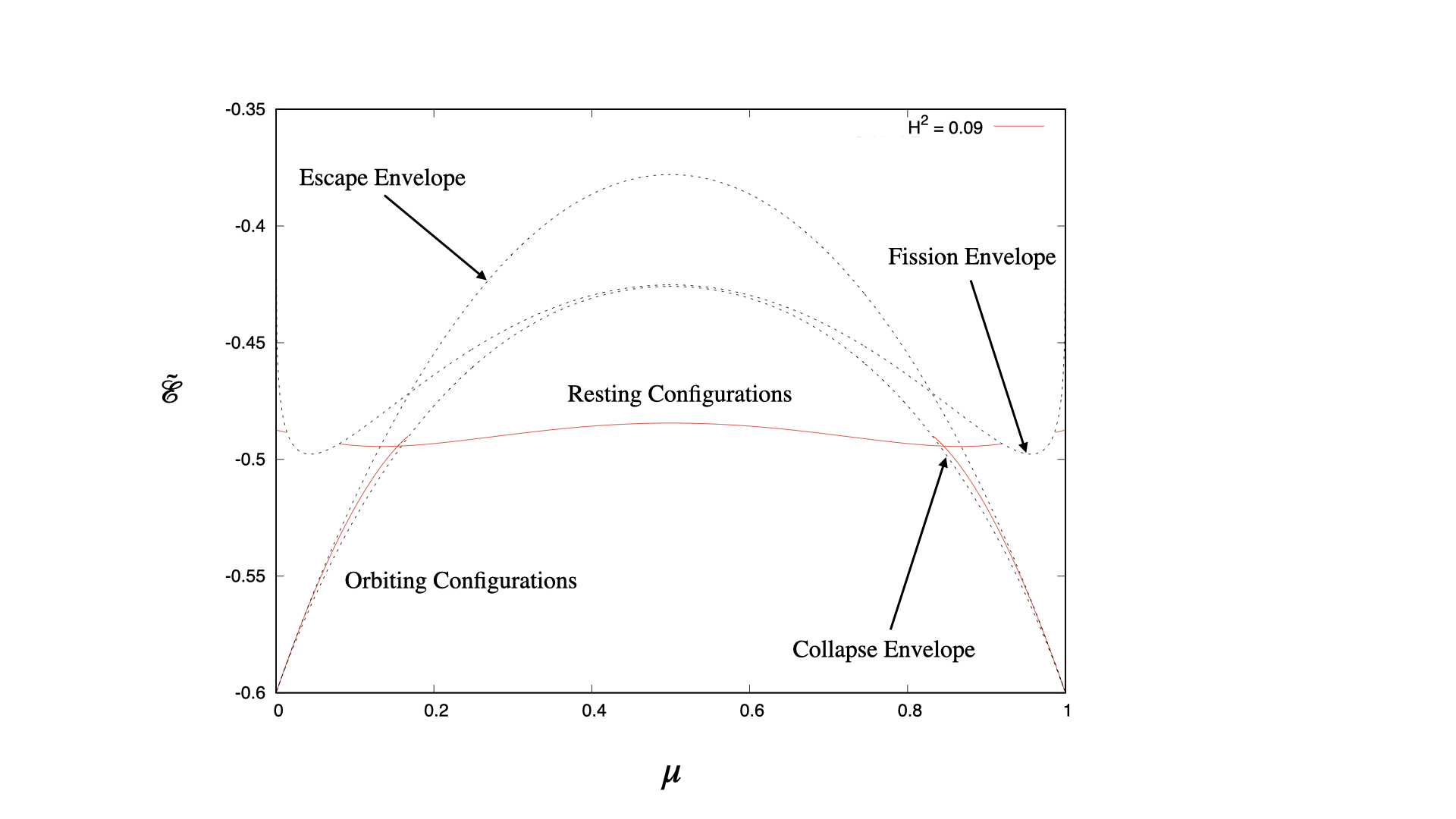}
    \caption{Detail of the $H^2 = 0.09$ energy phase diagram. {This example shows that both orbital or contact binary end states can co-exist at a given value of angular momentum, and even at the same value of mass ratio in some regions.}}
    \label{fig:ERO_09}
\end{figure}

The previously identified collapse and fission envelope curves play a key role in this diagram. When above the fission envelope, only orbiting equilibria can occur as a final minimum energy state. Thus, whenever a resting equilibrium line intersects with the fission envelope, it ceases to exist because the two bodies would undergo fission and enter orbit about each other. Conversely, when within the collapse equilibrium envelope, only resting equilibria can occur as a final minimum energy state. Thus, whenever an orbital equilibrium line intersects with the collapse envelope, it also ceases to exist and corresponds to a resting equilibrium. The resting and orbital lines for $H^2 = 0.09$ provide a clear example of both of these cases, and is shown in detail in Fig.\ \ref{fig:ERO_09}. We note that between the fission and collapse envelopes it is possible for both orbital and resting equilibria to exist at a given value of angular momentum and mass fraction. 

Within the region between the envelopes the energy phase diagram clearly shows the possibility of multiple equilibria for a given angular momentum at a given mass fraction. While there will always be a lower energy state that the system could seek out, we note that either of the equilibria will be stable. Thus, if the system initially has a larger energy, as it is dissipating energy over time, if the energy dissipation rate is slow, it may be more likely for the system to settle into the higher energy state, as it would encounter this first. Conversely, if energy dissipation is rapid, then it may jump over the higher energy state and settle in the minimum energy configuration. This was studied earlier for a simple, equal mass, 3-body system in \cite{gabriel2016energy}, where it was shown that as a function of system dissipation strength different outcomes would occur on average. 

\subsection{Energy and Angular Momentum Diagram}

\begin{figure}[!ht]
    \centering
    \includegraphics[width=0.98\linewidth]{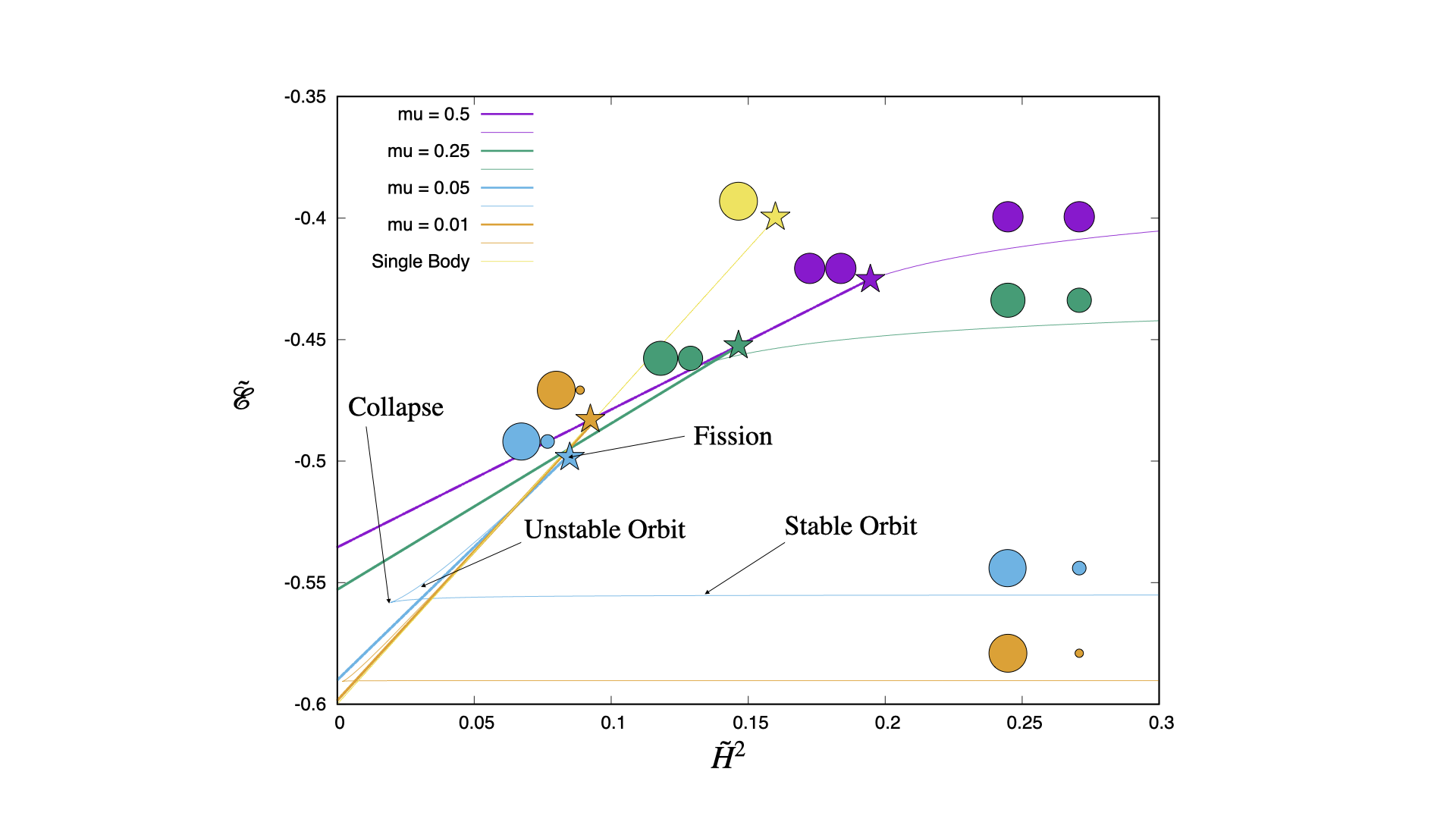}
    \caption{Energy-Angular Momentum chart of relative equilibria for the two-body problem for different mass fractions. Specific transition points and orbits are indicated for the curve at $\mu = 0.05$.}
    \label{fig:2BP_EvsH2}
\end{figure}

A third perspective on this phase diagram is shown in Fig. \ref{fig:2BP_EvsH2} in the energy and angular momentum space. This clearly shows the relationship between energy and angular momentum at fixed mass fractions. Here, for a given mass fraction, at zero angular momentum up to the fission value the energy increases linearly with $\tilde{H}^2$. As the angular momentum is increased the system will reach the fission value, where there will be two orbital equilibria which exist. One is unstable and lies at a local energy maximum, and disappears when the system reaches the fission angular momentum. The other is stable, and lies at a lower energy. Moving past the fission state, only the stable orbital equilibrium exists, and is defined for all larger angular momentum, with its energy asymptotically approaching the system self-potential energy as angular momentum increases. This form of the phase diagram has been studied previously for resting relative equilibria in the $N$ body problem \cite{scheeres2017constraints}.

\subsection{Full Phase Function}

\begin{figure}[!ht]
    \centering
    \includegraphics[width=0.98\linewidth]{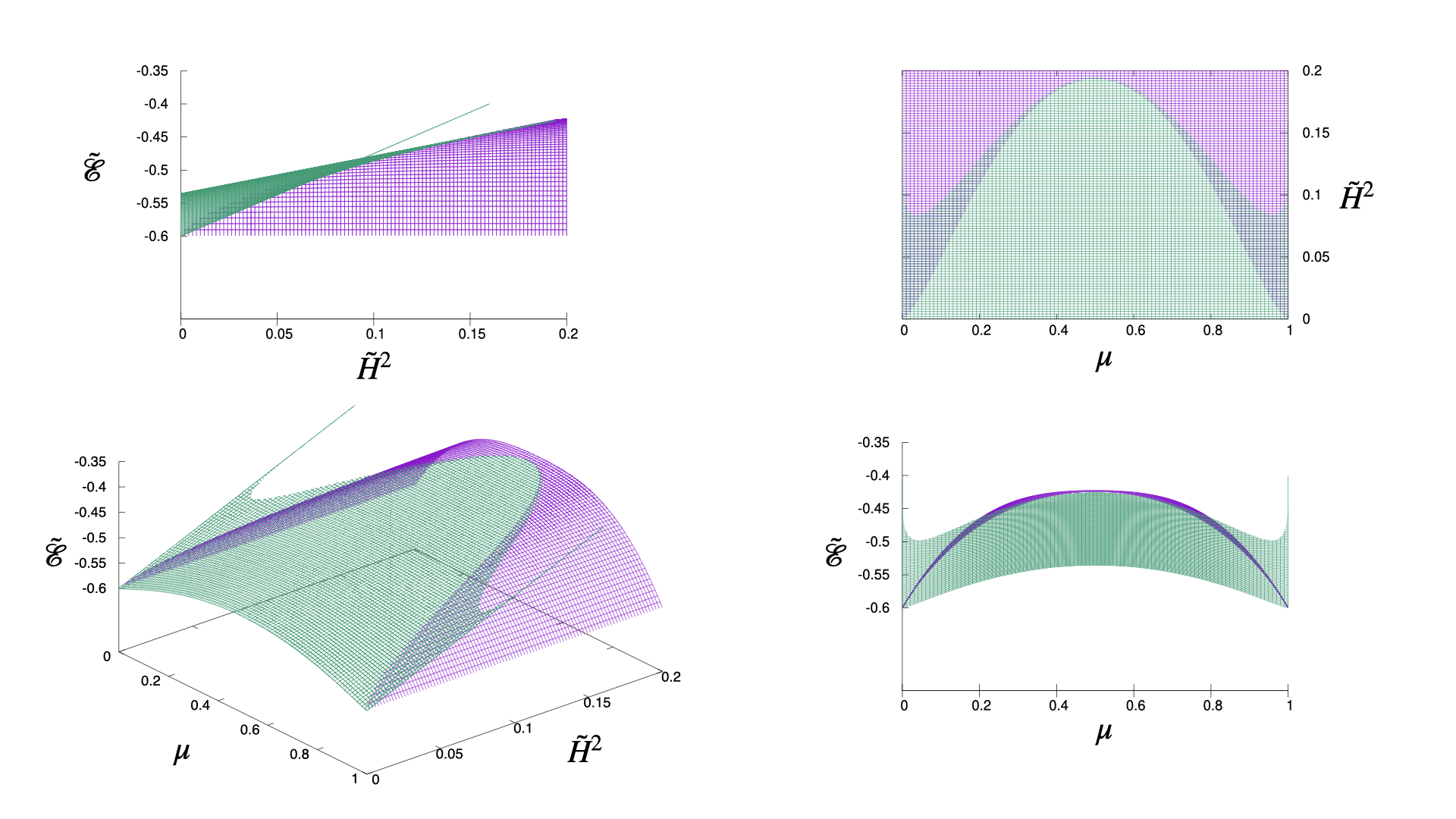}
    \caption{Full 3-dimensional view of the Phase Diagram showing the resting equilibria (green) and the orbiting equilibria (purple). {Also shown are the projections of the phase function, which can be compared with the previous plots.} }
    \label{fig:phase_3d}
\end{figure}

In Fig.\ \ref{fig:phase_3d} we show a three-dimensional view of this phase diagram, which has only been viewed in cross-section in the previous figures. 
This perspective clearly shows the different families of equilibrium solutions and the transitions between them. Every point in this space is a minimum energy configuration. Points between the fission and collapse surfaces can have multiple equilibrium states, while points outside these surfaces will either only have minimum orbiting or resting equilibria when all excess energy is removed from the system. 

\section{Non-Equilibrium Excited Configurations}



The previous phase diagrams are ideal in that they assume that all excess energy has been dissipated in the system. While many systems in nature may come close to this ideal, many do not. In particular we find a clear difference in rubble pile asteroids between those that have settled into a single or contact binary configuration and those that have settled into an orbital binary configuration in terms of how far these systems are found from their minimum energy state. 
In the following we introduce the ways in which these systems can be excited, and how that can be represented on the phase diagram. 

\subsection{Contact Binary Excitation}

Most single and contact binary asteroid systems are found in a minimum energy state of uniform rotation about their maximum moment of inertia, exceptions to this usually have a very low angular momentum (which has been shown to lead to very long dissipation timescales \cite{burns_safronov, harris_tumbling}). 
{When the components of a contact binary come into contact, excess energy can be dissipated efficiently due to the surface forces of friction and coefficient of restitution acting within the mass distribution. If the system angular momentum is low enough so that the given configuration can exist as a contact binary system, then as the excess energy is dissipated the system can remain in contact. Eventually, the only excess energy in a stable contact binary system would be in its excited spin state, which is then dissipated due to tidal flexure \cite{burns_safronov, harris_tumbling}. 
For our contact binary model it is possible to develop simple constraints on an excited contact binary asteroid in terms of its spin limits to exist as a cohesionless rubble pile and spin limits to exist as a contact binary. 
For our simple system these situations can be evaluated exactly. }
%
The ideal contact binary body of two spheres resting on each other is a prolate body and has its minimum moment of inertia through the contact point between the bodies, equal to the sum of the inertias of the two spheres, or $I_m = I_S(\mu)$. The maximum moments of inertia are symmetric about the minimum moment of inertia axis, and are equal to the sphere moments of inertia plus the moment due to the displacement of the centers of mass by the sum of the radii, or $I_M = I_S(\mu) + \mu(1-\mu)R(\mu)^2$. 

The total angular momentum of the system and the energy can be expressed as equal to $H = I_D \omega_l$ and ${ E} = \frac{1}{2} I_D \omega_l^2$, where $\omega_l$ and $I_D$ are the equivalent spin rate and moment of inertia of a simple body with the same total angular momentum and energy (see \cite{scheeres_asteroid_book} for a detailed description). In this form, since the angular momentum is constant, we express the energy as ${ E} = \frac{1}{2} \frac{H^2}{I_D}$. Then the dynamic inertia $I_D$ represents the level of excitement in the system. First, it must lie in the range $I_m \le I_D \le I_M$, so that when $I_D = I_M$ the system is in its lowest energy rotation state, rotating about its maximum moment of inertia. Conversely, the maximum energy rotation state is when $I_D = I_m$ and the system rotates about its minimum moment of inertia axis, and provides the maximum energy state that a rotational body can have. Thus the energy lies in the interval $\frac{1}{2}\frac{H^2}{I_M} \le { E} \le \frac{1}{2} \frac{H^2}{I_m}$. We note that for the minimum energy states discussed above the system was always assumed to spin about its maximum moment of inertia. 

As the spin energy is increased, by decreasing $I_D$, the body spins about both its minimum moment of inertia axis and about its maximum moment of inertia axis. The solutions can be found explicitly as \cite{scheeres_asteroid_book}
\begin{eqnarray}
	\omega^2_m & = & \frac{H^2}{I_m^2} \frac{\left(I_M/I_D - 1\right)}{\left(I_M/I_m-1\right)} \\
	\omega^2_M & = & \frac{H^2}{I_M^2} \frac{\left(1 - I_m/I_D \right)}{\left(1 - I_m/I_M\right)} 
\end{eqnarray}
The total angular rate of the body is then $\omega = \sqrt{ \omega^2_m + \omega^2_M }$. 

For the spin rate about the minimum moment of inertia axis, $\omega_m$, we must have $\omega_m \le 1$ to ensure that the rubble pile body can be stable and not disaggregate. This limit can be solved for as a limit on the dynamic moment of inertia, yielding 
\begin{eqnarray}
	\frac{1}{I_D} & \le & \frac{1}{I_M} \left[ 1 + \frac{I_m^2}{H^2} \left(\frac{I_M}{I_m} - 1\right)\right] 
\end{eqnarray}
We note that $I_D$ must always be greater than $I_m$, providing an upper limit of $1/I_m$ on the above inequality. For small $H^2$, the bounding limit passes this upper limit, meaning that at most the small body can spin about its minimum moment of inertia. For large $H^2$, the bounding limit will approach spin about the maximum moment of inertia, meaning that the excited body cannot spin solely about its minimum moment of inertia and can only be excited by a limited amount before spin about the minimum axis passes the disaggregation threshold. 

For spin rates about the maximum moment of inertia, $\omega_M$, we need to consider whether the body can fission due to centripetal acceleration exceeding gravitational attraction. This limit can be expressed as $\omega_M^2 \le \frac{1}{R^3(\mu)}$, which we note will always be less than the disaggregation spin rate. This can also be solved as a limitation on the dynamic moment of inertia 
\begin{eqnarray}
	\frac{1}{I_D} & \ge &  \frac{1}{I_m} \left[ 1 - \frac{I_M^2}{H^2 R^3} \left(1 - \frac{I_m}{I_M}\right)\right] 
\end{eqnarray}
Analysis of this limit shows us that if, at a given value of $H^2$ and $\mu$, a contact binary resting configuration is possible, then as the system is excited rotationally,  the excited system will not undergo fission as with increasing excitation its spin rate about this axis decreases. If, instead, the combination of $H^2$ and $\mu$ lies outside of the resting envelope, then at higher energy excitations it may be possible for the system to stay in contact, however as it dissipates energy there is an energy at which it will fission and, if the mass ratio stays fixed, it must seek out an orbiting minimum energy configuration. This is an interesting transition, and means that an excited body can sometimes exist as a contact binary even though it must undergo fission as it dissipates energy. We note that this discussion completely disregards the geophysical limits on failure within a rubble pile asteroid in favor of these unambiguous mathematical limits. This additional analysis is left as future work. 

\begin{figure}[!ht]
    \centering
    \includegraphics[width=0.75\linewidth]{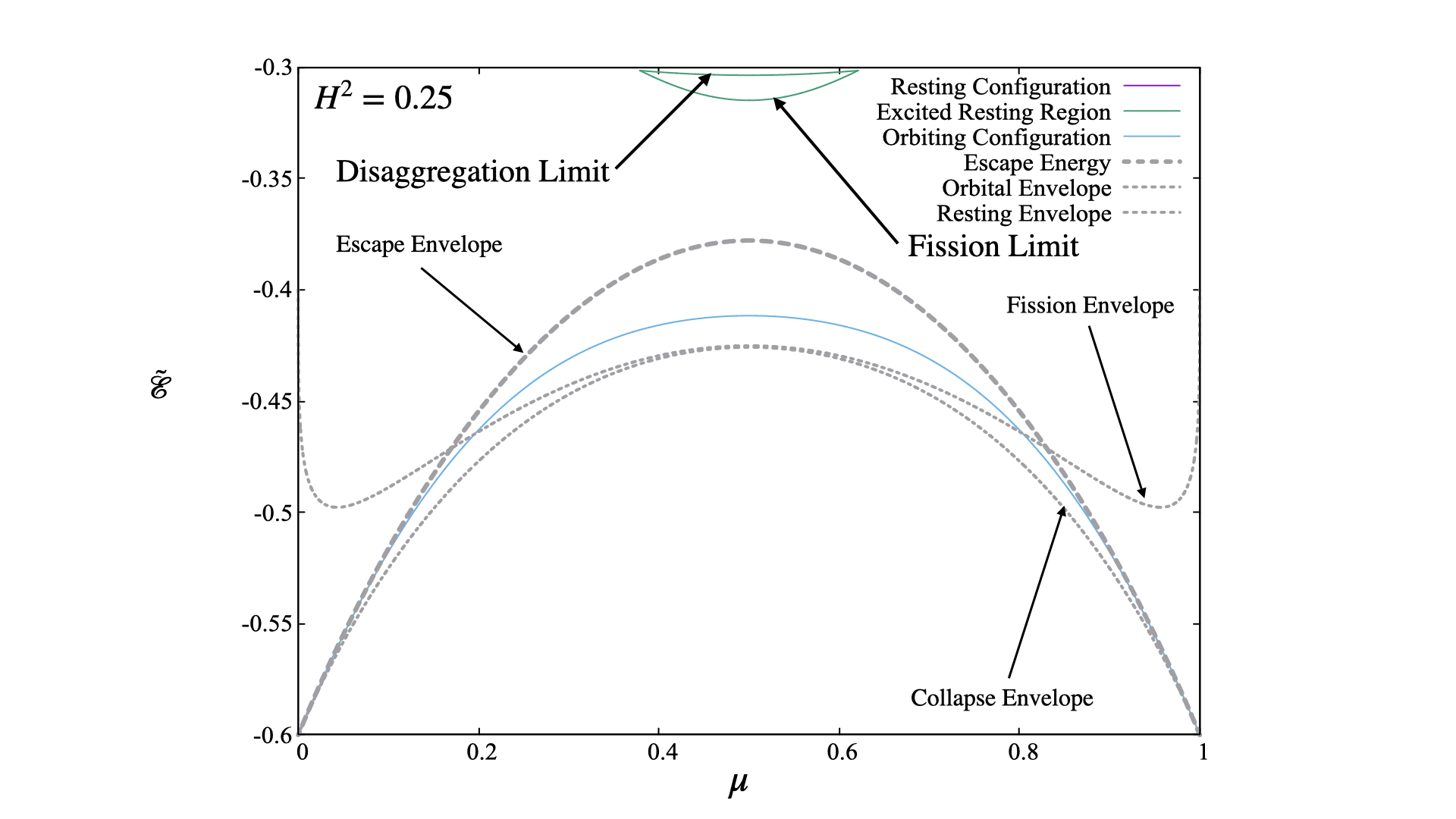}
    \includegraphics[width=0.75\linewidth]{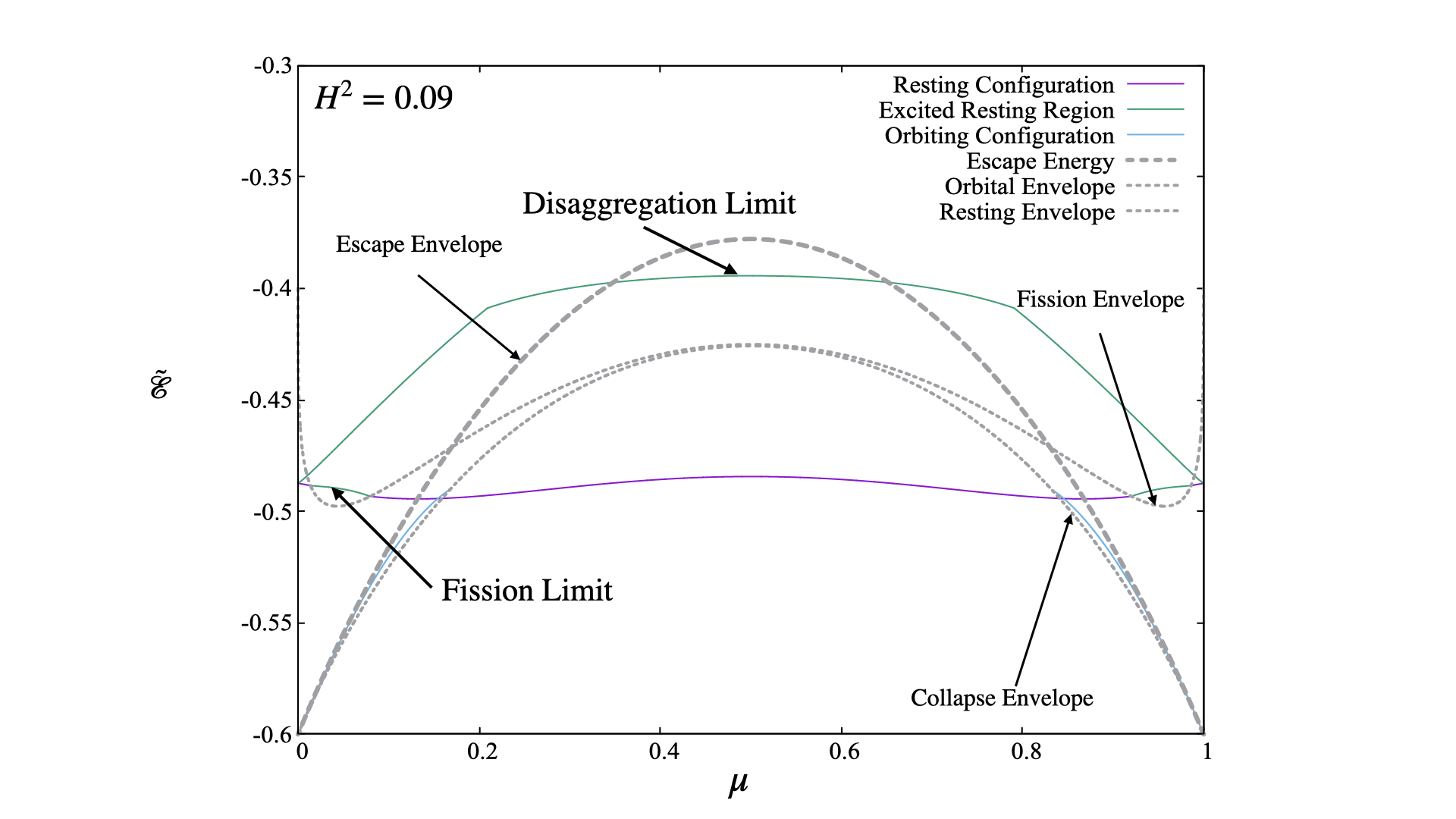}
    \includegraphics[width=0.75\linewidth]{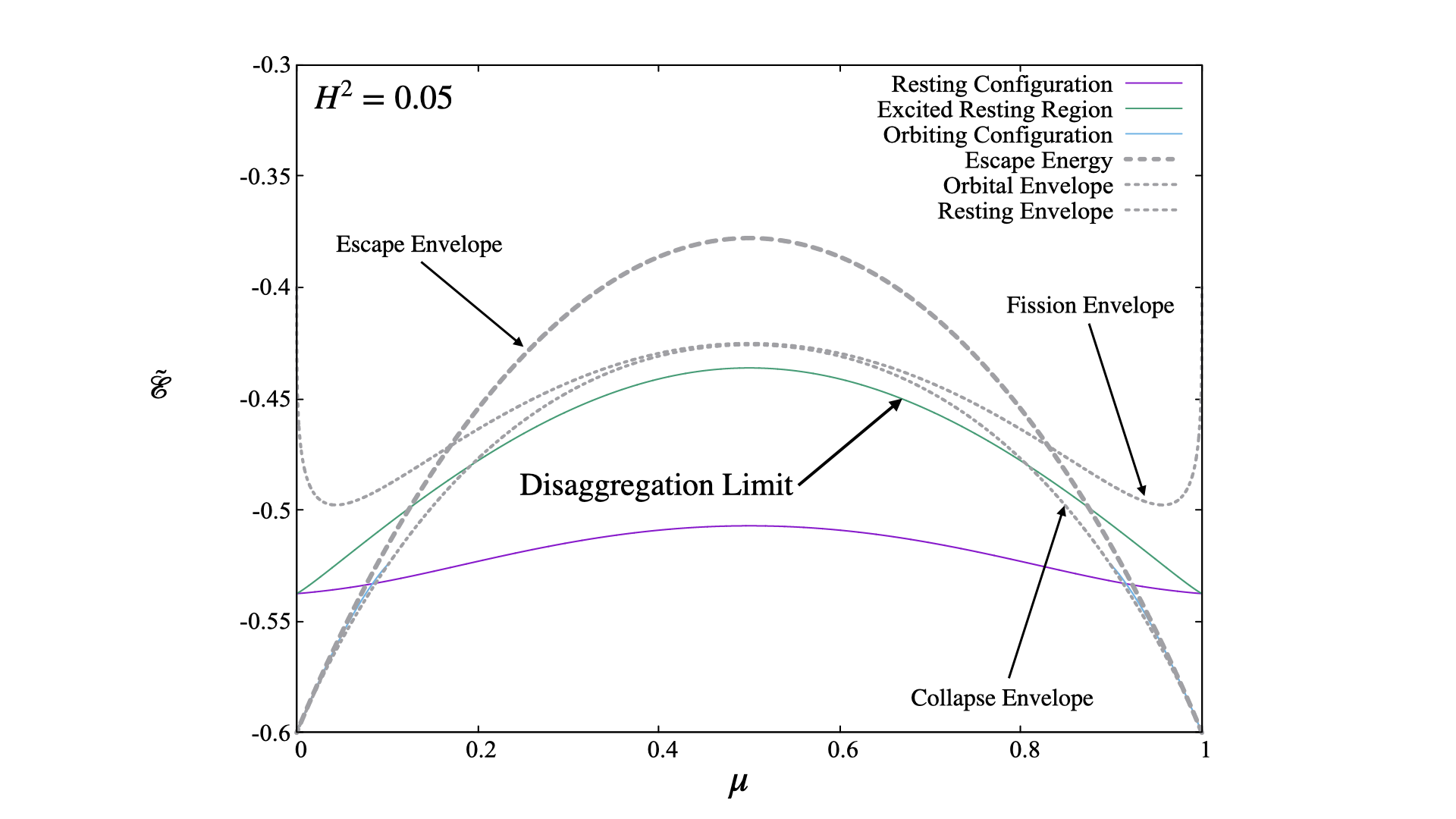}
    \caption{Spin excitation limits for the three cases $\tilde{H}^2 = 0.25, 0.09, 0.05$. }
    \label{fig:spin_energyCB250905}
\end{figure}


In Fig. \ref{fig:spin_energyCB250905} we show these excited spin state limits at different levels of angular momentum. For a high level of angular momentum ($\tilde{H}^2 = 0.25$), when there are no resting solutions, it is still possible for an excited contact binary body to spin at less than the disaggregation limit (about its minimum moment of inertia) and less than its fission limit (about its maximum moment of inertia). The plot shows this narrow region where a tumbling body could be stable. However, any body spinning within that region would eventually fission under energy dissipation. In this case, the body would then be in an orbital configuration and, if it continues to dissipate energy, would eventually settle into a stable, doubly synchronous solution. We note that the fissioned system would have sufficient energy to escape, however.
In the figure we also show more moderate levels of angular momentum ($\tilde{H}^2 = 0.09$), when there are both resting and orbiting solutions present. Here we see that there is only a small region of excited rotational motion that will fission when it dissipates energy. This corresponds to the range where there are no contact binary solutions at this angular momentum. At every other point in the excited domain the system would dissipate to a resting contact binary asteroid. Above the upper curve the system would disaggregate. 
We also show a low level of angular momentum ($\tilde{H}^2 = 0.05$), when there are no orbiting solutions and only resting solutions. Here, all excited contact binary bodies would eventually dissipate into a stable contact binary body rotating about its maximum moment of inertia.



\subsection{Orbital Binary Excitation}

In contrast to contact binary bodies, orbital systems are usually not found in overall minimum-energy configurations. {The minimum energy state for a binary is to be in a doubly synchronous state.  While there are binaries that have been observed in this state, not many of them have been found -- possibly due to detection bias against finding them.}  Instead, the majority of observed binary asteroids are in an excited singly-synchronous state, with the secondary synchronized with the orbit but the primary spinning faster than the orbit rate. The final minimum energy state of such a situation (ignoring non-gravitational forces) is for tides to dissipate excess energy and drive the system to a larger orbit radius while the body spin rates slow, ideally leading to a doubly-synchronous state. For systems with a large angular momentum, however, the final orbital distance between the bodies will be large, making the system susceptible to disruption due to 3rd body perturbations, non-gravitational effects, or impacts. 
Another way in which binary systems can be in an excited state is for the orbit itself to be eccentric. While there are binary systems with eccentric orbits, they are not as common as the singly-synchrounous, circular binaries. Predictions are also that the eccentricity is more rapidly dissipated than the excess spin rate \cite{meyer2023energy}. 

To evaluate the ways in which binary asteroid orbits can be excited, we focus on the limiting behavior. Indeed, while there is only one minimum energy orbiting state for a given angular momentum and mass ratio, there are many ways in which such a state can be excited. For these simple systems there are three key terms that will control the excitation level of a binary system: the spin rate of the components, the distance between the components, and the eccentricity of the orbit. Let's consider each of these in turn. For simplicity we assume that both components are spinning at the same rate, $\tilde{\omega}$, as then the angular momentum contribution and energy contribution of these terms are just $I_S(\mu) \tilde{\omega}$ and $I_S(\mu) \tilde{\omega}^2 / 2$, respectively. The more general case where the bodies spin at different rates can be evaluated, but is left for future work. 

\begin{figure}[!ht]
    \centering
    \includegraphics[width=0.75\linewidth]{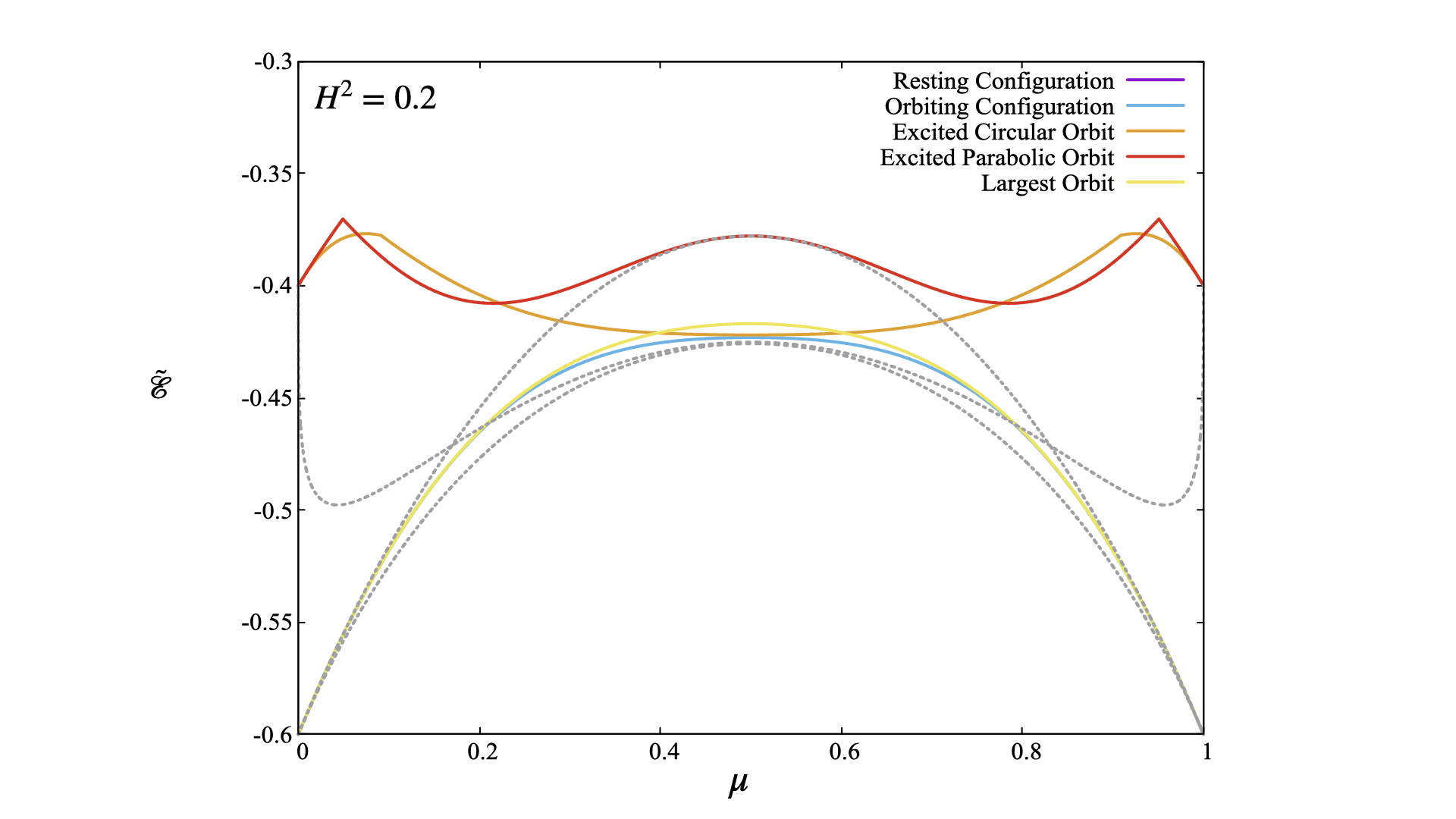}\\
    \includegraphics[width=0.75\linewidth]{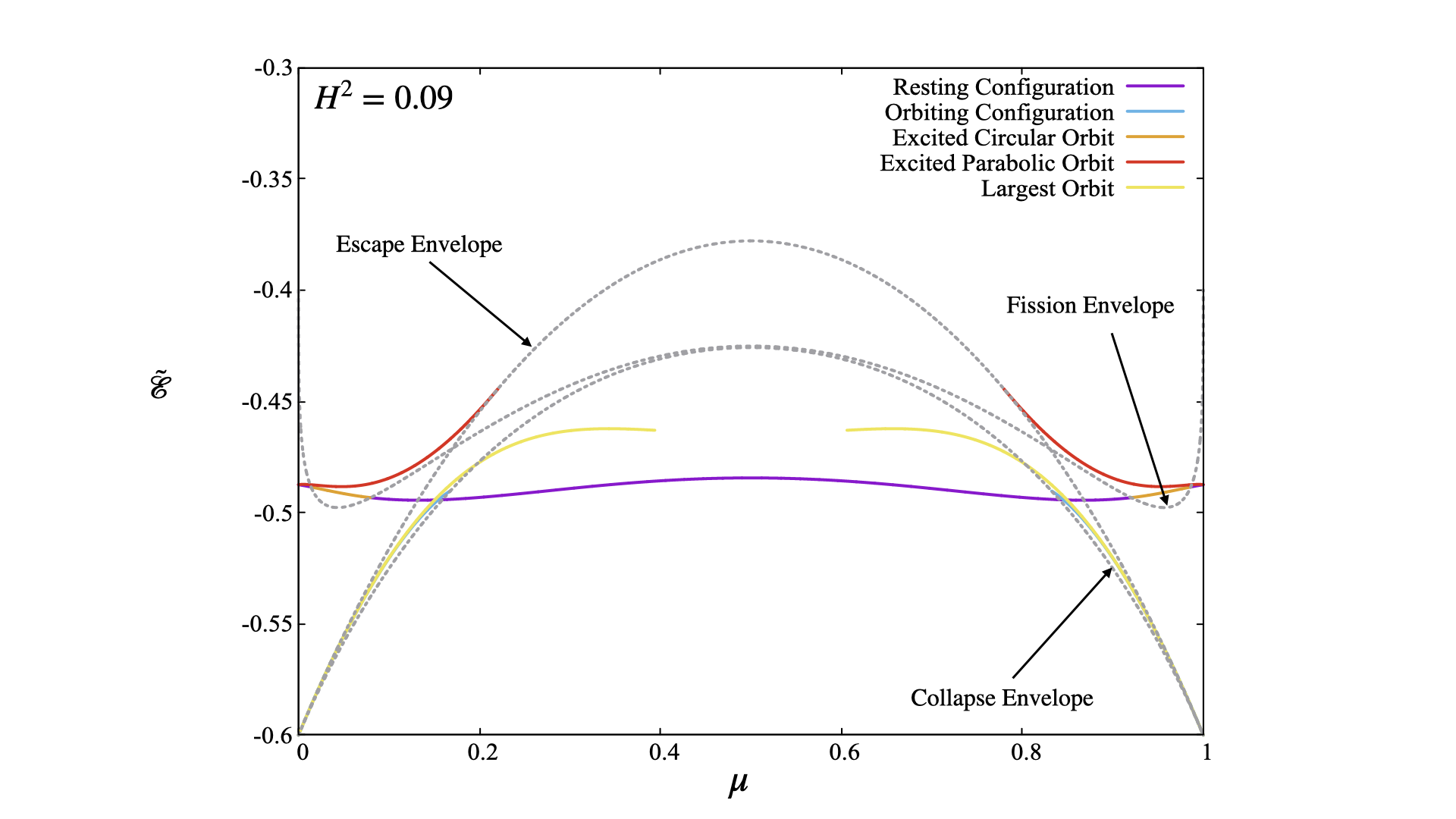}\\
    \includegraphics[width=0.75\linewidth]{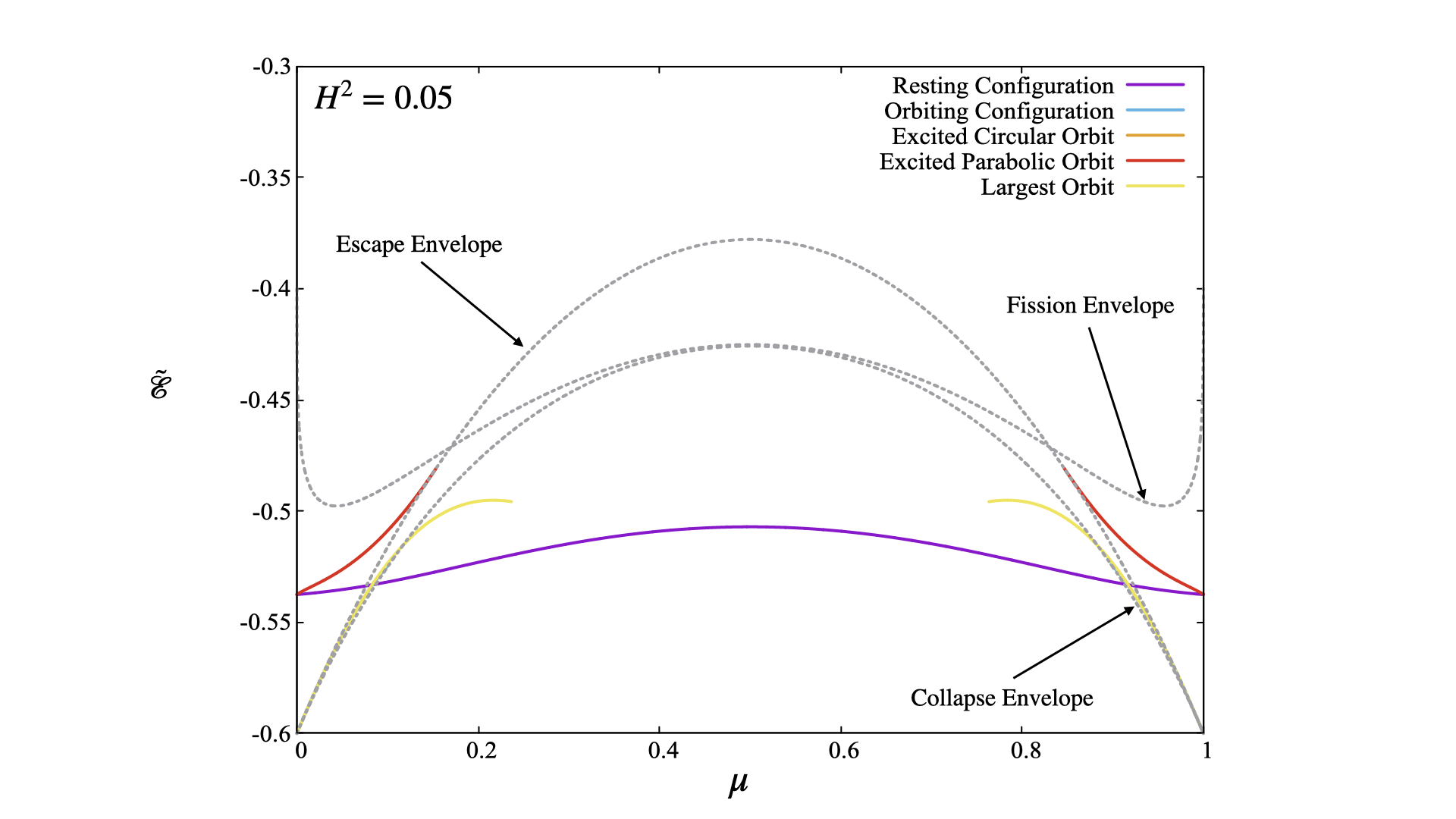}
    \caption{Orbit excitation limits for $\tilde{H}^2 = 0.2, 0.09, 0.05$. }
    \label{fig:energyORB200905}
\end{figure}

The minimum distance between the components can be equated to the periapsis radius of the orbit, $\tilde{q}$, and must lie in the interval $R_S(\mu) \le \tilde{q} < \infty$. Finally, the eccentricity must lie in the limit $0 \le e \le 1$. Given the bound on the periapsis, the situation when $e=1$ corresponds to a parabolic orbit. Given the combined periapsis and eccentricity, the orbital angular momentum and energy are then, $\mu(1-\mu)\sqrt{\tilde{q}(1+e)}$ and $-\frac{\mu(1-\mu) (1-e)}{ 2\tilde{q}}$, respectively. As each of these parameters are varied, however, we must keep the angular momentum fixed. Thus we have the system of equations with free variables $\tilde\omega$, $\tilde{q}$ and $e$, 
\begin{eqnarray}
	\tilde{H} & = & I_S(\mu)\tilde{\omega} + \mu(1-\mu)\sqrt{\tilde{q}(1+e)} \\
	\tilde{E} & = & \frac{1}{2}I_S(\mu)\tilde{\omega}^2 - \frac{\mu(1-\mu)(1-e)}{2\tilde{q}} - \frac{3}{2}I_S(\mu) 
\end{eqnarray}
For a fixed angular momentum we can then fix one of our variables and vary the other two. From our derivation, we note that $\tilde{E} \ge \tilde{\cal E}_{orb}$ at every level of angular momentum and mass ratio. In the following we explore a few different types of excited orbits. These are all represented in Fig.\ \ref{fig:energyORB200905} for angular momentum values of $\tilde{H}^2 = 0.2, 0.09, 0.05$ to show a range of different energy excitation limits, as described below. Note that we only assume that the bodies spin in the same direction as the orbit. It is possible to get more extreme energy excitation by allowing the bodies to spin in the opposite direction of the orbit, which would allow both of these to be larger for a fixed angular momentum. This extreme situation (which has not been seen in nature for rubble pile binary asteroids) could be explored in future analysis. 

\paragraph{Excited Circular Orbits}
First, let $e = 0$, so $\tilde{q}$ is the semi-major axis and can theoretically vary from contact to infinity. Denote the equilibrium distance value as $\tilde{d}'$, then we first consider $\tilde{q} < \tilde{d}'$. As the orbit shrinks, the bodies must spin faster to keep the angular momentum constant. There are two possible limits that can occur, either $\tilde\omega = 1$, at which the two bodies will disaggregate, or $\tilde{q} = R_S(\mu)$, at which point the bodies will touch but the spins will not be synchronized with the orbit, leading the system to rapidly dissipate energy (and potentially change the mass ratios). In Fig. \ref{fig:energyORB200905} this is represented as the Circular Excited orbit case (orange line). The kink in the curve is where the limit switches from $\tilde{\omega} = 1, \tilde{q} > R_S(\mu)$ (towards the $\mu = 0,1$ sides) to $\tilde{\omega} < 1, \tilde{q} = R_S(\mu)$ (in the inner region). 

We can also consider larger circular orbits, $\tilde{q} > \tilde{d}'$, which would cause the spin rate to decrease. Taking the limit at $\tilde{\omega} = 0$ we then get the Largest Circular Orbit case (yellow line), which we see stays close to the orbital equilibrium solution. The full range of excited circular orbits go from the yellow curve, touching the equilibria at the equilibrium distance, and then expand for growing energy up to the orange curves. We ignore cases where the bodies are spinning in the opposite direction of the orbital angular momentum -- which should allow for a further increase in energy. While such a situation may occur, we postpone analyzing this for the future. 

\paragraph{Excited Parabolic Orbits}
We can also consider the other eccentricity limit of $e=1$ and evaluate how much the orbit energy can be boosted while still keeping the system bound. Here the orbit energy is 0 and the angular momentum is again constrained. As the periapsis decreases the bodies spin faster, with the same constraints as occurs for the circular orbits, with the bodies either spinning at the disaggregation limit or the periapsis reaching the contact limit. The Excited Parabolic Orbit (red line) denotes this limit, again with a transition at the kink in the curve. 
We can also increase the periapsis radius away from the orbital radius, up to the zero spin case. At this limit they intersect with the escape energy limit. 

%

The case $\tilde{H}^2 = 0.25$ exhibits all of these excited orbits. For lower angular momenta, where $\tilde{H}^2 = 0.09$, there is a small interval of excited circular orbits but most of the bounding occurs due to the excited parabolic and the larger circular orbits. In  the case where $\tilde{H}^2 = 0.05$ only these latter limits are possible for such a low level of angular momentum. 

\subsection{Combined Excitation Example}

\begin{figure}[!ht]
    \centering
    \includegraphics[width=0.98\linewidth]{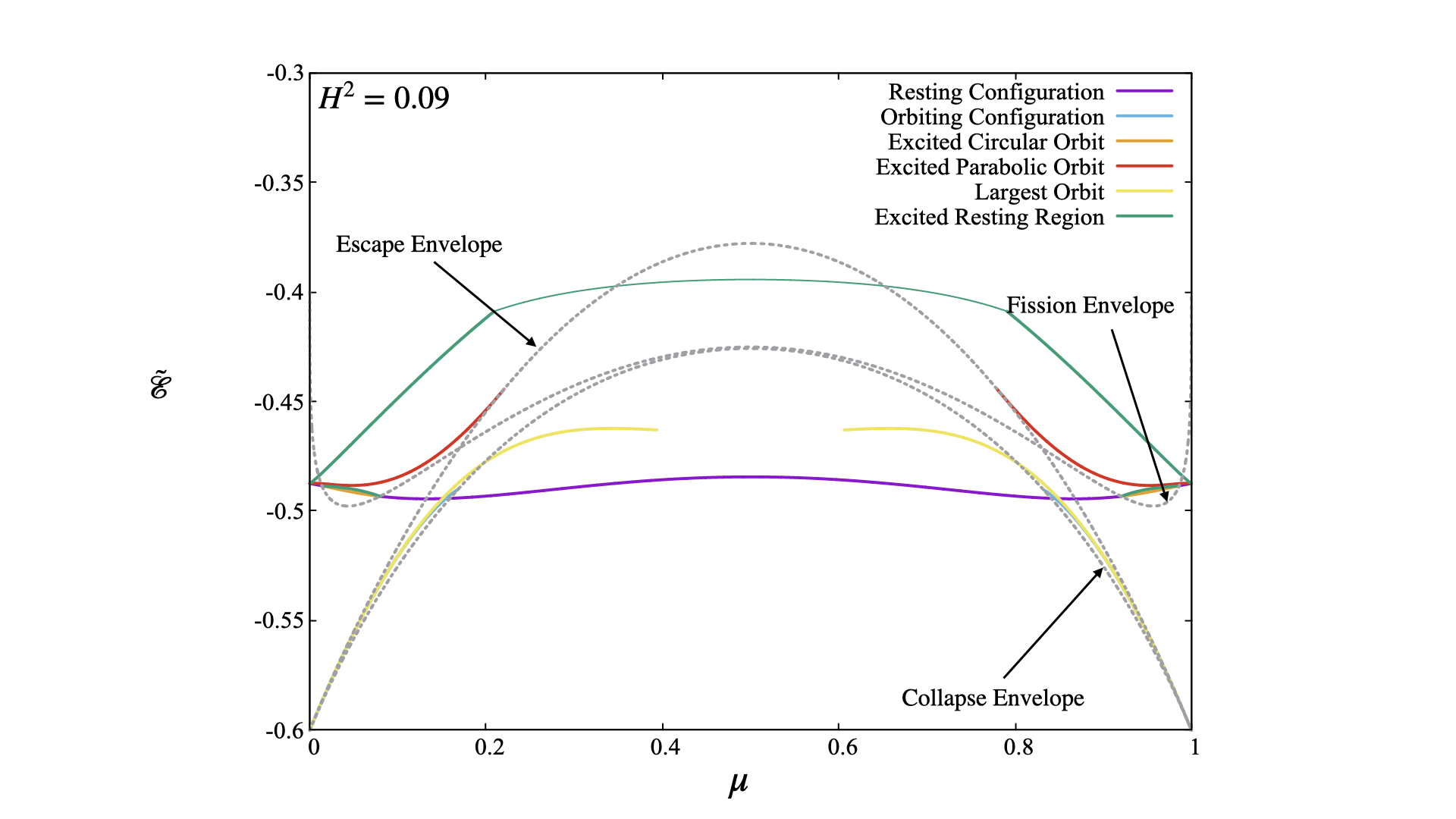}
    \caption{Combined Orbit and Resting excitation limits for $\tilde{H}^2 = 0.09$. }
    \label{fig:energyORB09_total}
\end{figure}

To finish this section we briefly consider all of the different excitation states. In Fig.\ \ref{fig:energyORB09_total} all of the different orbital and contact binary excitation cases are shown for the case $\tilde{H}^2 = 0.09$. This combined case is the more realistic situation, as under excitation any of these non-equilibrium states (or combinations of them) could be achieved across the range of system morphology. 

\section{Example Cases}

To connect the phase function to physically relevant systems, we present a first principles computation and map a number of asteroid systems into the model. First we set up a simple system that can mimic the original mass distribution envisioned in the problem motivation and develop predicted ranges on initial density for different morphologies. Following this we look at a number of specific asteroid binaries, contact binaries, single asteroids and asteroid pairs to better understand how these systems map into the phase diagram. 

\subsection{Formation Model}

\begin{figure}[!ht]
    \centering
    \includegraphics[width=0.98\linewidth]{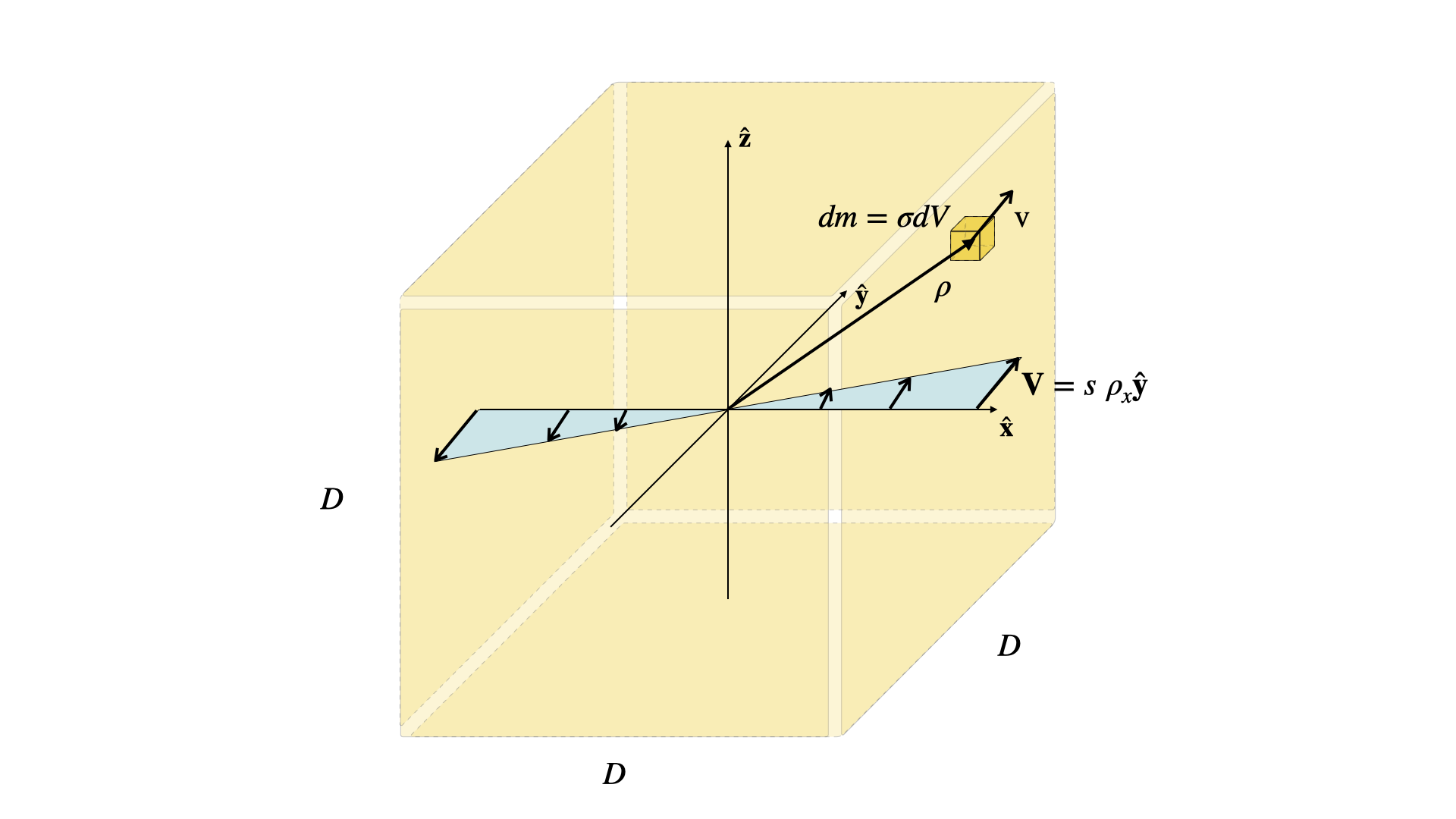}
    \caption{Simple distribution of material with a velocity shear across it. }
    \label{fig:distribution}
\end{figure}

We first connect the above model to a physical formation and collapse scenario. Consider an ideal distribution of particles that may mimic a post-collision situation. Assume a total mass $M$ of non-rotating bodies with grain density $\sigma_G$ are uniformly distributed over a cubic region with dimension $D$ on each side, making an overall bulk density of $\sigma = M / V$ (see Fig.\ \ref{fig:distribution}).

A mass element within this distribution is located by the vector ${\bf \rho} = \rho_x\hat{\bf x} + \rho_y\hat{\bf y} + \rho_z \hat{\bf z}$, with the center of mass of the distribution being at the origin.  The bodies all share a common velocity, plus a gradient in the velocity along the $\hat{\bf x}$ direction, giving the volume a net, non-zero angular momentum. The velocity of any point in the distribution is then given as ${\bf v} = {\bf V}_C + s \rho_x \hat{\bf y}$. The parameter $s$ is the velocity shear, and means that particles will have linearly changing velocities as they move across the volume, which will give the distribution a non-zero angular momentum. Note that the center of mass ${\bf R}_C$ is at the cube's center and that the velocity of the center of mass is ${\bf V}_C$, as the average of the shear is zero. The total mass of this distribution is $M = \int_{\cal B} \sigma dV = \sigma D^3$, where ${\cal B}$ designates the distribution, $\sigma$ is the initial density of the distribution and $dV = d\rho_x d\rho_y d\rho_z$ is a differential volume term. For modeling the gravitational attraction and energy, we assume the system can be modeled as a sphere with total mass $M$ (note that the expected error in the gravitational attraction as compared to a cube is on the order 10-50\% across the surface of the cubic region \cite{werner_scheeres_poly}). 

For the particles to not immediately escape (although they may escape later through gravitational interactions), the shear is initially limited so that the outermost particles have less than escape speed
\begin{eqnarray}
	\rho s & < & \sqrt{ 2 {\cal G} M / \rho } 
\end{eqnarray}
If we assume that the total mass of the distribution less than $\rho$ is $M = \sigma (2\rho)^3$, then a limit on the shear to prevent immediate escape is
\begin{eqnarray}
	s & < & 4 \sqrt{{\cal G}\sigma} 
\end{eqnarray}
Define the dimensionless shear as $\tilde{s} = s / \sqrt{{\cal G}\sigma}$, giving the initial limit $\tilde{s} < 4$. 


\paragraph{Bulk Parameters of the Distribution}
Given this simple distribution the total angular momentum, kinetic energy and gravitational potential energy of the distribution relative to the center of mass can be computed, yielding 
\begin{eqnarray}
	{\bf H} & = & \int_{\cal B} {\bf \rho}\times\left({\bf v}-{\bf V}_C\right) \sigma dV = \frac{s \ \sigma}{12} D^5 \hat{\bf z} \\
	T & = & \frac{1}{2} \int_{\cal B} \left({\bf v} - {\bf V}_C\right)\cdot\left({\bf v} - {\bf V}_C\right) \sigma dV = \frac{s^2 \ \sigma}{24} D^5 \\
	{\cal U} & = & \int_{\cal B}\int_{\cal B} \frac{dm \ dm'}{\left|{\bf \rho} - {\bf \rho}'\right|} \ \sim \ - \frac{3}{5} \frac{{\cal G} M^2}{(D/2)} = - \frac{6}{5} {\cal G} \sigma^2 D^5
\end{eqnarray}
where for convenience we use the self potential of a constant density sphere for the total gravitational potential of the system. Using the moment of inertia of a constant density cube, the moment of inertia about the angular momentum vector direction is
\begin{eqnarray}
	I_H & = & \frac{1}{6} \sigma D^5 
\end{eqnarray}

Introducing the non-dimensionalizations defined previously, the normalized values of angular momentum and energy are
\begin{eqnarray}
	\tilde{H}^2 & = & \frac{1}{144} \tilde{s}^2 \tilde{D}^4 \label{eqn:H2norm} \\
	\tilde{T} & = & \frac{1}{24} \tilde{s}^2 \tilde{D}^2 \\
	\tilde{\cal U} & = & - \frac{6}{5 \ \tilde{D}} \\
	\tilde{I}_H & = & \frac{1}{6}\tilde{D}^2 
\end{eqnarray}
Then the total energy of the system is $\tilde{E} = \tilde{T} + \tilde{\cal U}$ and the amended potential is $\tilde{\cal E} = \frac{\tilde{H}^2}{2\tilde{I}_H} + \tilde{\cal U}$:
\begin{eqnarray}
	\tilde{E} & = & \frac{\tilde{s}^2\tilde{D}^2}{24} \left[ 1 - \frac{144}{5 \tilde{s}^2 \tilde{D}^3}\right] \\
	\tilde{\cal E} & = & \frac{\tilde{s}^2\tilde{D}^2}{48} \left[ 1 - \frac{288}{5 \tilde{s}^2 \tilde{D}^3}\right] \\
	\tilde{E} - \tilde{\cal E} & = & \frac{\tilde{s}^2\tilde{D}^2}{48}
\end{eqnarray}
The initial excess energy ($\tilde{E}-\tilde{\cal E}$) in the system equals one half the initial kinetic energy and is also equal to the initial kinetic energy bound up in the angular momentum term. 

A necessary condition for all the masses to condense is that the total energy be negative, $\tilde{E} < 0$, which constrains the shear further to:
\begin{eqnarray}
	\tilde{s} & < & \frac{12}{\sqrt{5}} \frac{1}{\tilde{D}^{3/2}} 
\end{eqnarray}
This is not a sufficient condition for total collapse, but does further restrict the initial shear across the body. If the energy is positive, that does not mean that the entire distribution can escape, but it does mean that at least one particle must escape. Thus, if the energy is negative it is possible for the system to condense without losing additional mass. 
Let us directly compare the shear to the zero energy value, defining the factor 
\begin{eqnarray}
	f & = & \tilde{s} / \left(  \frac{12}{\sqrt{5}} \frac{1}{\tilde{D}^{3/2}} \right)
\end{eqnarray}
Substituting into Eqn.\ \ref{eqn:H2norm} yields
\begin{eqnarray}
	\tilde{H}^2 & = & \frac{1}{5} f^2 \tilde{D} \label{eq:H2dist}
\end{eqnarray}
{which relates the normalized size and shear of the distribution to its initial angular momentum.}

\paragraph{Limiting Effect of the Sun}

We should also consider the effect of the sun to see what limits this provides on the initial size of the volume to be considered. In dimensional units the Hill Sphere is approximated as
\begin{eqnarray}
	d_H & = & \left(\frac{M}{3M_S}\right)^{1/3} a
\end{eqnarray}
where $M$ is the total mass of the collapsing region, $M_S \sim 2\times10^{30}$ kg is the mass of the sun, and $a$ is the semi-major axis of the center of mass, where 1 astronomical unit $\sim 1.5\times10^{11}$ m. Normalizing this by the unit radius of the condensed mass, and setting $M = 4\pi\sigma_B R^3 / 3$ gives the dimensionless Hill radius
\begin{eqnarray}
	\tilde{d}_H & = & \left(\frac{1}{3}\right)^{1/3} \left[\frac{4\pi\sigma_B}{3 M_S}\right]^{1/3} a
\end{eqnarray}
Substituting a density of $2500$ kg/m$^3$ and evaluating the distance at 1 AU gives 
\begin{eqnarray}
	\tilde{d}_H & \sim & 180 A
\end{eqnarray}
where $A$ is the asteroid-sun distance in AU. Thus, in the asteroid belt from 2.2 to 3.2 AU the normalized distance ranges from 400 to almost 600. If the initial size of the volume is set to be equal to the Hill sphere, the size compression of the condensed mass will be on the order of 500. 

\paragraph{Dimensional Values}

Now the angular momentum in Eqn.\ \ref{eq:H2dist} is compared to the theoretical limits found in this paper. For example, for the system to have any resting configurations possible, it must have $\tilde{H}^2 < 0.2$, providing a constraint on the normalized shear and overall compaction of material of $f \sqrt{\tilde{D}}  <  1$. 
If we take values of $\tilde{D}$  on the order of $100 \rightarrow 500$, then for a contact binary outcome to be possible requires $f = 0.1 \rightarrow 4.5\times10^{-2}$, or the normalized shear should vary over $\tilde{s} = 5.4\times10^{-4} \rightarrow 2.1\times10^{-5}$. 

These values can be mapped back to metric units given a bulk density $\sigma_B$. Assuming an S-Type asteroid, take $\sigma_B = 2500$ kg/m$^3$.  Then conservation of mass gives the equality $\sigma D^3 = 4\pi/3 \sigma_B R^3$, or $\sigma = 4\pi/3 \sigma_B / \tilde{D}^3$. Thus, for the range of $\tilde{D}$ the range of initial densities are $\sigma = 1\times10^{-2} \rightarrow 8\times10^{-5}$ kg/m$^3$. By scaling the initial shear velocity and heliocentric semi-major axis of the distribution these values can be adjusted to a wider range of possibilities. 
These results are just given to enable numerical simulations to be mapped into the phase function. 

\subsection{Comparison with a Sampling of Asteroids}

In addition to the focus on the initial conditions for the formation of a rubble pile system, it is also useful to use the results as a prism through which to consider current asteroid systems. To do this we take a sampling of single asteroids, contact binary asteroids, orbital binary asteroids and asteroid pairs and appropriately represent them on the phase function. 
All of these bodies have surely been subjected to evolutionary effects that have changed their total angular momentum and energy. So by mapping them into the diagram it is also possible to constrain the evolution of a given system. Table \ref{tab:1} summarizes the systems and the corresponding normalized angular momentum and energies of single, contact binary and orbital binary asteroids, while Table \ref{tab:pair} lists some asteroid pairs. For all of these asteroids more detailed information is given in the Appendix. Figures \ref{fig:H2Data} and \ref{fig:EData} show these systems represented on the angular momentum and energy phase diagram. The different asteroid types are discussed in turn. 

\begin{figure}[!ht]
    \centering
    \includegraphics[width=0.98\linewidth]{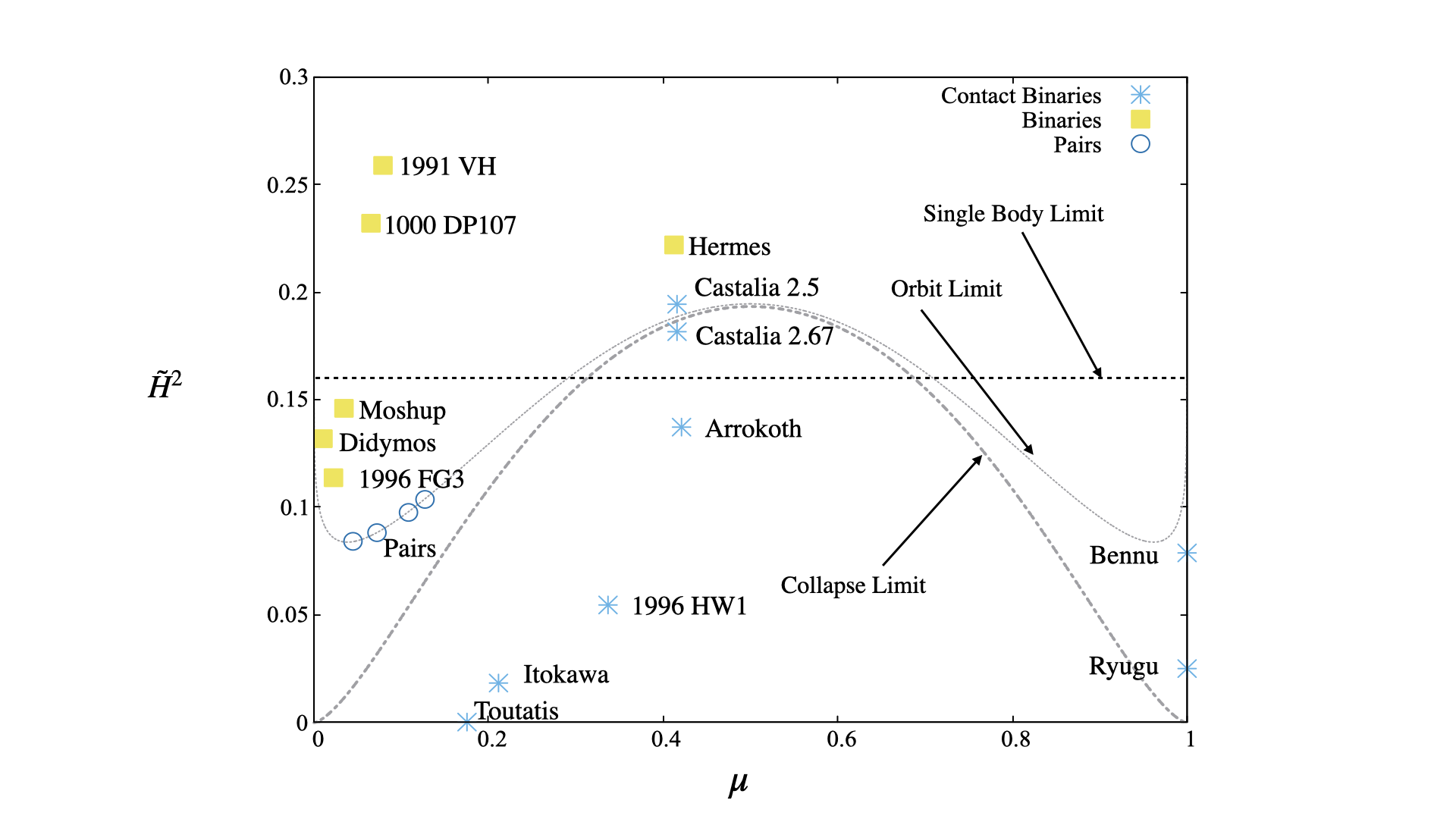}
    \caption{Binary and Contact Binary bodies projected into the $\tilde{H}^2$-$\mu$ space. }
    \label{fig:H2Data}
\end{figure}

\begin{figure}[!ht]
    \centering
    \includegraphics[width=0.98\linewidth]{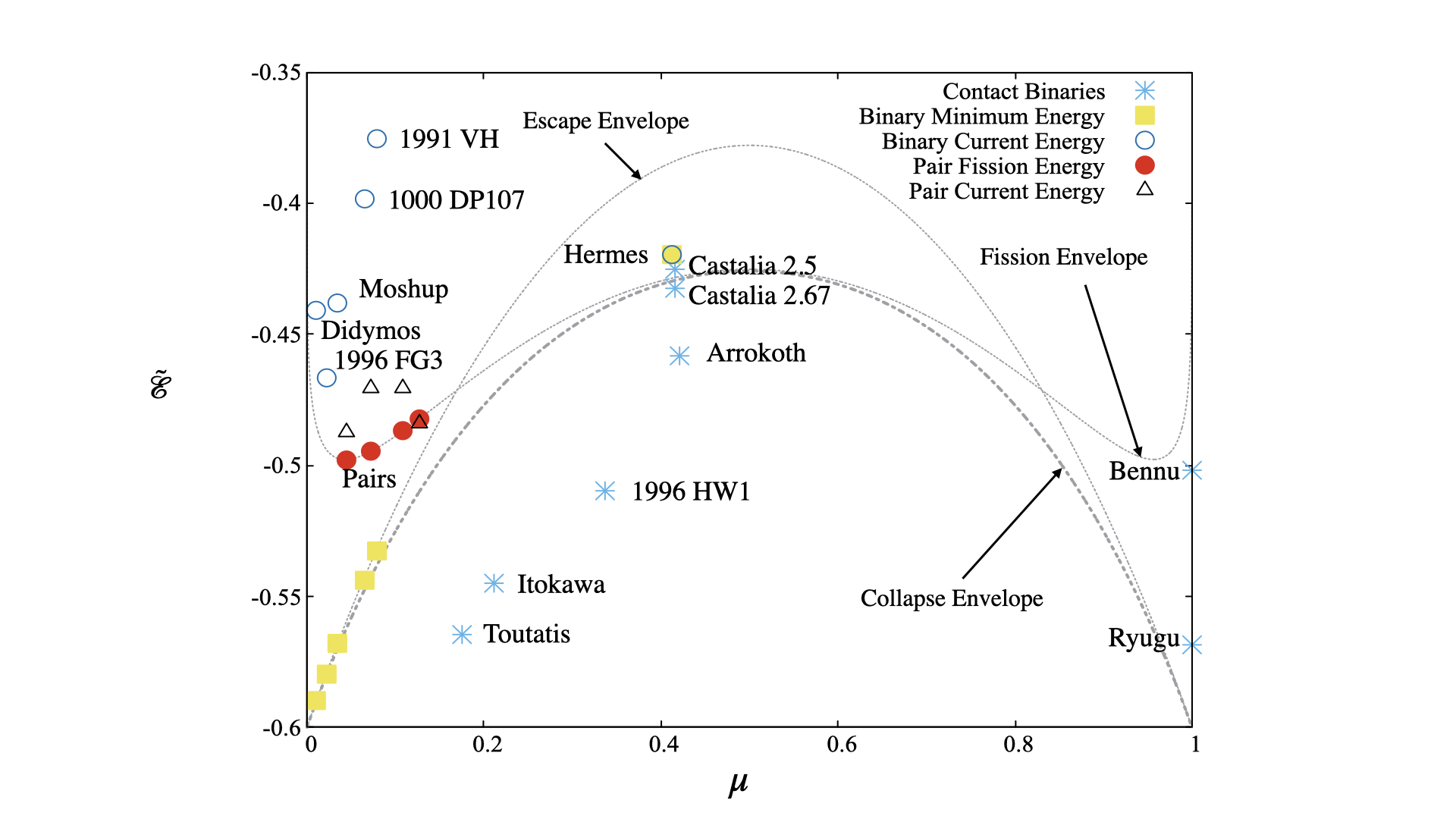}
    \caption{Binary and Contact Binary bodies projected into the $\tilde{\cal E}$-$\mu$ space. }
    \label{fig:EData}
\end{figure}

\paragraph{Single Asteroids:} On the phase function single asteroids appear as bodies on the $\mu = 1$ line, which in our simple model is defined solely by their normalized spin which is mapped into an angular momentum and energy. Here we note that both asteroids so mapped, Bennu and Ryugu, lie in the stability region of these plots, as expected. We do note that there are known single asteroids that spin beyond the stability limit as defined here, and which have been hypothesized to have cohesive strength \cite{toshi_1950DA}. Despite this, the vast majority of asteroids greater than a few hundred meters would lie within the stability bound along the $\mu = 1$ line. 

\paragraph{Contact Binary Asteroids:} The contact binary asteroids all lie within the collapse envelope (although see the discussion on Castalia below). We note that the lowest energy and angular momentum asteroid is Toutatis, which is in an excited spin state yet which has such a low overall angular momentum that energy excitation is not visible on the graph. The highest energy and angular momentum bodies in the selection are Arrokoth and Castalia, which are both near equal mass contact binaries. 

The asteroid Castalia shows an additional application of this diagram, that it can be used to test hypotheses on the densities of bodies. 
In \cite{meyer_ApJL} an assumed density for Castalia is given as 2.5 g/cm$^3$, which is consistent with its S-Type status. However, when plotted on the angular momentum diagram we see that this assumed density places the body into the orbital regime. While this could be due to uncertainties in our model, or this body having some level of cohesive strength, it could also be due to a poor density assumption. In fact, a small change in the body's density to 2.67 g/cm$^3$ moves the body into the contact binary regime. Also, in the energy diagram, at this level of normalized angular momentum it can be shown that it lies in a consistent spot, but very near to a fission rate for this body. 

\begin{figure}[!ht]
    \centering
    \includegraphics[width=0.98\linewidth]{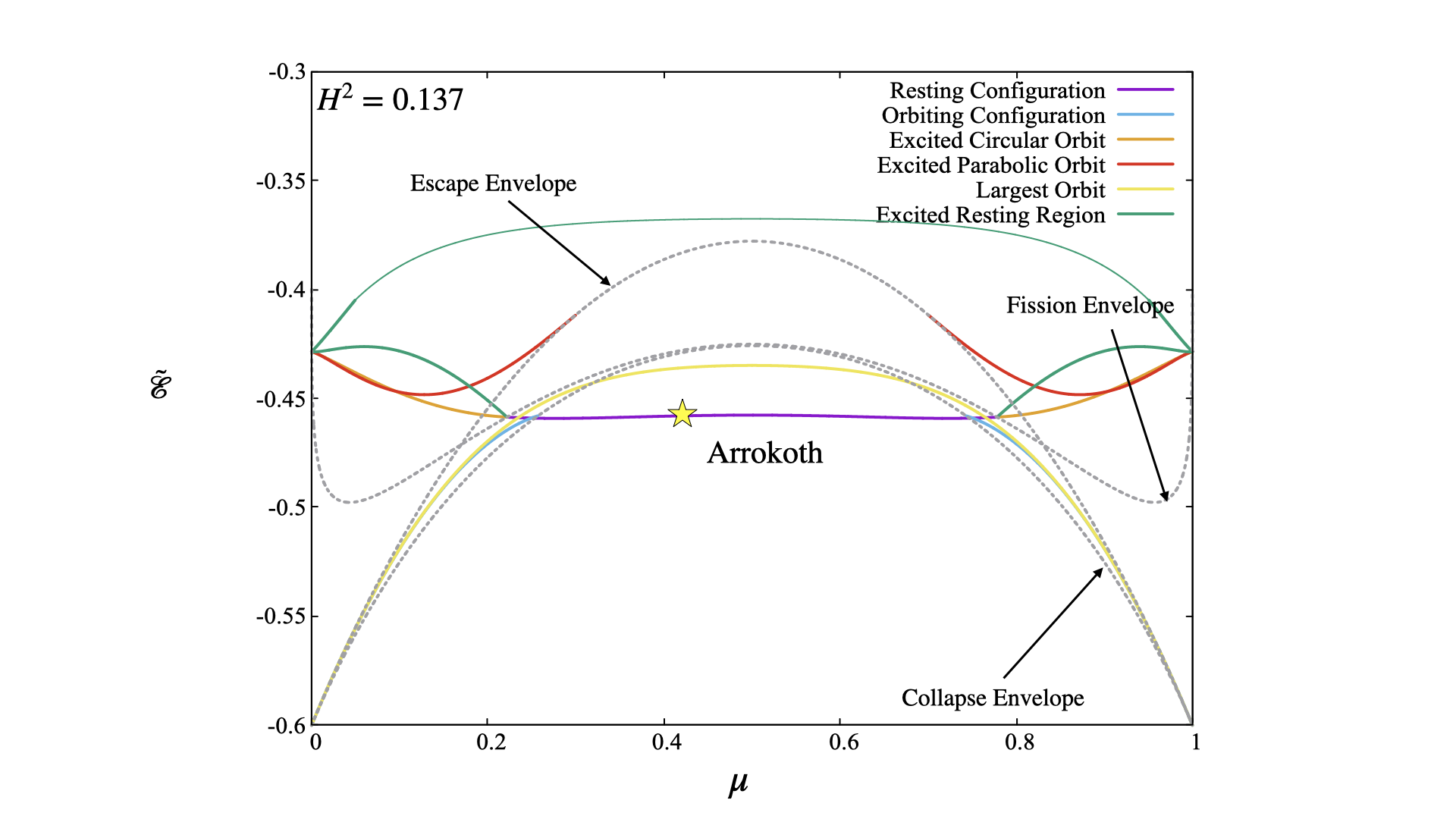}
    \caption{Possible excited states of Arrokoth given its angular momentum. }
    \label{fig:arrokoth}
\end{figure}

The Kuiper Belt Object Arrokoth has also been mapped into this diagram as a contact binary, consistent with its morphology \cite{spencer2020geology}. An interesting exercise for this body is to show the possible energy excitation states for its estimated angular momentum. This is relevant as there is significant debate regarding the formation of this body. Figure \ref{fig:arrokoth} shows the different excited states possible for Arrokoth at its given angular momentum. Here we see that both orbiting and tumbling solutions are allowed, which would dissipate energy into this final state through tidal flexure. Utilizing the phase diagram in this way opens up the number of possible hypotheses to be considered for this body. 

\paragraph{Orbital Binary Asteroids:} 

The orbital binary asteroid examples generally show a large separation between their minimum energy states and their current state, consistent with the earlier discussion. The only exception is the doubly synchronous asteroid Hermes, which also has a mass fraction close to 0.5. A fundamental question is whether the orbital binaries show the natural evolution from a formation state or whether they have been strongly perturbed by exogenous effects. The most highly excited binary is 1991 VH, whose secondary is in a tumbling state and which has documented chaotic dynamics. Recent research has shown that this system is consistent with having been perturbed by a close Earth flyby in the past \cite{meyer2024earth}. Another example is the binary asteroid 1996 FG3 which has been determined to lie in a Tide-Binary YORP equilibrium state \cite{scheirich_FG3}, which implies exogenous forces are at work. For a binary system in a BYORP equilibrium state, the BYORP effect is subtracting orbital angular momentum, shrinking the orbit, which is being offset by the orbit expansion due to the tidal angular momentum transfer from the primary to the orbit. Thus, as the net angular momentum of the system should be decreasing, this should cause the system to migrate to a lower angular momentum value. However, the YORP effect is also likely operating on the primary, which could offset this -- especially if the Tangential YORP effect is driving the primary spin rate to increase. It has even been hypothesized that such a combination of effects could result in a system equilibrium \cite{golubov_DSBYORP}. 

Finally, there have been many papers published focusing on the possible spin-fission formation of the Didymos binary asteroid system. Here a relevant question is how that system would migrate within these diagrams as these scenarios play out. Here we note that both the binaries Moshup and Didymos are relatively close to the stability spin limit of a single asteroid, which would be expected for a body that has undergone spin-fission. However, since both of these bodies have total angular momentum less than the single body angular momentum limit it does imply that the angular momentum of the system has either undergone some evolution since the body underwent spin fission or that angular momentum was shed by escaping components following its spin fission. 

\paragraph{Asteroid pairs:} 
It is more difficult to represent asteroid pairs on these diagrams. What we know are the relative size (and implied relative mass) of the two bodies and the spin rate of the primary at least. For some select asteroid pairs we also know the spin period of the secondary and know the asteroid Type. In this paper we identified all S-Type asteroid pairs in \cite{pravec2019asteroid} for which we had the spin periods of both bodies. With this information we can infer the fission angular momentum and energy (just as a function of the mass fraction $\mu$), while with the assumed density (from the S-Type determination) and spin periods we can compute the current energy. As the total angular momentum results from a vector sum it is not useful to just add these scalar values. In the future, if the spin poles of these bodies are known, it may be possible to also constrain the current angular momentum of the system. The results are consistent with the conclusions from \cite{pravec_fission, Pravec_clusters} that these asteroid pairs are consistent with originally being a contact binary asteroid that subsequently fissioned and escaped. For this to occur, the progenitor contact binary body must lie outside of the escape energy envelope, which they all do. From a simple observation of the asteroid pairs on the energy diagram (Fig.\ \ref{fig:EData}) we note that the current energies for three of the pairs are elevated from the hypothetical fission value. Given the large uncertainties in the various assumptions and models, the results are consistent with the theory adopted in \cite{pravec_fission}. The fourth body is distinguished by the energy difference being near zero. We do not try to draw any conclusions from these representations and observations, but do note that this theory allows these asteroid pairs to be viewed as part of a single system. 


\begin{table}[h]
\caption{Asteroid systems mapped in terms of total angular momentum ($\tilde{H}^2$), current energy ($\tilde{E}$) and minimum energy ($\tilde{\cal E}$). Details on the data for these computations are given in Table \ref{tab:ast_data} in the Appendix.}

\begin{center}
\begin{tabular}{ | c | c || c | c | c | c | }
\hline
Asteroid & Morphology & Mass  & & & \\
 &  &  Fraction & & & \\
 Name & Type & $\mu$ & $\tilde{H}^2$ & $\tilde{ E}$  & $\tilde{\cal E}$ \\
\hline
\hline
Bennu & Single& 1 & 0.07862164 &-0.50172295 &  -0.50172295 \\
\hline
Ryugu & Single & 1 & 0.02516818 & -0.56853977 & -0.56853977 \\
\hline 
\hline 
Toutatis & CB & 0.176 & 0.00005716 & -0.56447682 & -0.56447682 \\
\hline
Itokawa & CB & 0.212 & 0.01829304 & -0.54514464 & -0.54514464\\
\hline
1996 HW1 & CB & 0.336 & 0.05457037 & -0.50951685 & -0.50951685 \\
\hline
Castalia (2.67) & CB & 0.416 & 0.18188452 & -0.43221363 & -0.43221363 \\
\hline
Castalia (2.5) & CB & 0.416 & 0.19425267 & -0.42505818 & -0.42505818 \\
\hline
\hline
Didymos & OB & 0.01088 & 0.13189348 & -0.44084728 & -0.589696 \\
\hline
1996 FG3 & OB & 0.02143 & 0.11402499 & -0.46639179 & -0.580123 \\
\hline
Moshup & OB & 0.03475 & 0.14619937 & -0.43806114 & -0.568031 \\
\hline
2000 DP107 & OB & 0.06454 & 0.23178631 & -0.39814168 & -0.543343 \\
\hline
1991 VH & OB &  0.07919 &  0.25902807 & -0.37535706 & -0.532510 \\
\hline
Hermes & OB & 0.412 & 0.22200373 & -0.41969119 & -0.41969119  \\
\hline
\hline
\end{tabular}

\end{center}
\label{tab:1}
\end{table}

\begin{table}[h]
\caption{Asteroid pairs in terms of predicted fission angular momentum ($\tilde{H}_F^2$) and energy ($\tilde{E}_F$), current energy ($\tilde{E}$) and difference in energies. Details on the data for these computations are given in Table \ref{tab:pair_data} in the Appendix.}

\begin{center}
\begin{tabular}{ | c | c || c | c | c | c | }
\hline
Asteroid  & Mass  & Fission & Fission & Current & Energy \\
 Pair & Fraction & AM & Energy & Energy  &  Difference \\
 Names & $\mu$ & $\tilde{H}_F^2$ & $\tilde{ E}_F$  & $\tilde{ E}$ & $\tilde{E}-\tilde{E}_F$ \\
\hline
\hline
Moore-Sitterly		& 0.04448734 & 0.08402131 & -0.49775656 & -0.48706450 &	0.01069206 \\
1999 RP27 & & & & & \\
\hline
Rheinland		& 0.10776534 & 0.09777121 & -0.48702487 & -0.47062228 &	0.01640259 \\
Kurpfalz & & & & & \\
\hline
1999 TU95		& 0.07203316 & 0.08823097 & -0.49474801 & -0.47071793 &	0.02403008 \\
2001 DO37 & & & & & \\
\hline
2001 QH293		& 0.12628958 & 0.10363697 & -0.48241482 & -0.48367947 &	-0.00126465 \\
2000 EE85 & & & & & \\
\hline
\end{tabular}

\end{center}
\label{tab:pair}
\end{table}

\clearpage

\section{Discussion}


This theoretical analysis develops conditions under which a collapsing rubble pile can form into a single or multi-body asteroid. The analysis is idealized in many ways, yet points out general constraints that will control the initial formation of a rubble pile asteroid and its subsequent evolution. Such a detailed perspective has not been analyzed analytically in this way previously in the asteroid literature. 
The driving question of this analysis is whether specific binary aspects of a collapsing rubble pile can be present at the initial stages of formation, or whether these are due to evolutionary effects. 
As a function of the normalized angular momentum in the collapsing system we find a range of different possible binary outcomes discriminated by angular momentum. 

For a large enough total angular momentum in a collapsing system, generally above $\tilde{H}^2 > 0.2$,  it can be impossible for a single body to form (see Fig. \ref{fig:phase2}). Above this limit, binary asteroids at any mass fraction can absorb the excess angular momentum in the orbit, with more equal mass fractions having a smaller orbit and higher energy. 
Even though the system cannot end up in a resting contact binary,  an excited binary may still be able to undergo impact with itself, which can cause reshaping of the components. Whenever such an impact between bodies occurs, it should rapidly shed energy due to their proximity or impact, and conceivably could easily transfer mass between the bodies, thus changing the mass ratio as well. 

For low values of angular momentum, $\tilde{H}^2 < 0.08$, the dominant minimum energy state will be a resting state (see Fig. \ref{fig:phase2}). We see that across all mass ratios this outcome is always possible. For a given angular momentum the energy of these resting states is always highest for near equal mass components. When considered as part of an evolving, dissipating system, this means that the first possible stable configurations to emerge from a collapsing distribution will be contact binaries with mass ratios near equal.  In \cite{nesvorny2010formation,nesvorny2021binary} it is noted that during gravitational collapse of KBOs, equal mass components are seen frequently. 
Thus, if the mass fraction $\mu$ is ``uniformly'' distributed during formation, then for mass fractions closer to 0.5 the amount of energy needed to be shed to achieve a stable configuration will be minimized, and thus over time it may be more likely for near equal mass contact binaries to condense out of the system first, creating a core for the remaining mass to accrete on. This could help explain the predominance of contact binary asteroids with similar size components. 

For values of angular momentum between these two limits there is not necessarily a unique final, stable state of the system. This implies that over a range of initial angular momentum values the final pathway of a system is not ordained, and that the detailed dynamics and interactions that occur during the dissipative stage of the system evolution can lead to different outcomes, even for the same mass ratio. This is interesting, as it injects a stochastic element into the model, one which would need to be tested using numerical simulations of appropriately designed situations (see \cite{gabriel2016energy} for such a study using a simple 3-body system). 

One interesting item to note in Fig. \ref{fig:phase2} is that for any angular momentum, the minimum energy binary orbit state will be one with a small mass fraction and the primary body spinning at a near-zero rate. 
The orbital size of these systems will be very large, as the small mass must be far from the primary in order to match the angular momentum. These systems will be very susceptible to disruption from third body effects. Also, it will be very difficult for these minimum energy orbits to form in finite time as the dissipation rate of energy will be very slow as the tidal flexure will be small and decreasing. The low value of energy also means that the prospective systems must shed significantly more energy to reach these configurations as compared to a contact binary with more equal mass ratio. Thus, it does not seem feasible that we find binary asteroids in this state, and in fact we note that all of the observed asteroids with lower mass fractions have elevated energies and are quite far from these minimum energy states. 
{Finally, we show that the phase diagram can also be used to constrain the past or future evolution of individual asteroids. When a given body is mapped into the diagrams, it is possible to evaluate the range of possible excitation states that the body may have had in the past, or to identify its future resting place if in an excited state. 
The diagrams can also be used to track how a body may evolve as its total angular momentum changes due to exogenous effects.  In order to make the diagram more useful in such characterizations, it will be important to develop a more sophisticated version of the model that incorporates non-spherical shapes. 
}



\subsection*{Extensions for Future Work}

There are several directions in which this current analysis could be extended. In the following a few of these are listed with some limited discussion. 

\paragraph{Non-spherical shapes:} It is possible to modify the spherical assumption in this analysis and replace the mutual and self gravitational potentials with more complex, non-spherical shapes. This was done earlier in \cite{scheeres_F2BP_planar}, and can change the symmetry in the mass fraction diagrams. These changes would also have to track the distorted geometry of these bodies, as fission and disaggregation spin rates would be expected to change given the different distances between body centers of mass. {We also note that if the bodies are allowed to deform, incorporating internal mechanics parameters such as friction \cite{holsapple_original}, then additional energy can be trapped in the body shapes, thus halting evolution at an even higher energy value.}

\paragraph{Inclusion of geophysical failure criteria:} Another modification could be focused on incorporation of more realistic, physically based geophysical failure criteria for the limits on the excited rotation and orbit states \cite{hirabayashi_survey}. These should, in general, further limit the range over which different morphologies could be stable, and could further constrain the original morphologies that condense out of the initial conditions. This also justifies the investigation of shapes that are not Jacobi or Maclaurin ellipsoids, as a body will require some internal friction or strength in order to maintain a non-equilibrium shape. 

\paragraph{Focus on large bodies: } The theory could be simplified to only track the creation of orbital binaries, assuming that all condensed bodies would reshape into single bodies. 
This would then more closely mimic the numerical simulations reported in \cite{nesvorny2010formation, nesvorny2021binary} and could provide insights into how energy and angular momentum will be redistributed in such systems. This analysis would ideally be performed using Maclaurin and Jacobi ellipsoids to more accurately track energy and angular momentum of the condensed bodies.  

\paragraph{Numerical simulations:} A fruitful area would be to use the phase diagram to motivate specific numerical simulations to test predictions. Such controlled runs could also supply necessary statistical data for further interpretation of formation probabilities. This could, in turn, lead to a more rigorous statement of what the expected formation probabilities of different morphologies could be.

\section{Conclusions}

An idealized analysis of the gravitational collapse of a rubble pile asteroid is derived as a function of angular momentum and mass fraction. An energy phase function is found that defines the conditions for the bodies to end up as a single body, an orbiting binary or a contact binary system. The phase function tracks the relative energy of different stable states, and could be used to determine the probability of different outcomes occurring. This function is also used to develop limits on how excited different systems can be before they would disaggregate. These diagrams also have a dual usage in terms of tracking the evolution of asteroid systems when subjected to exogenous effects that can change their angular momentum and energy. A call for numerical experiments that can inform the relative probabilities of different outcomes is made. Future directions for this work are also given. 

\paragraph{Acknowledgements}
This research was performed under the support of NASA grants from the YORPD program: 80NSSC24K0445 and 80NSSC22K0240. 

\section*{Appendix}
\subsection*{Explicit minimum energy orbital distance}

The orbital equilibrium condition given in Eq.\ \ref{eq:H2eq} can be rewritten as a polynomial for the relative distance $\tilde{d}$
\begin{eqnarray}
	\left[\mu(1-\mu)\right]^2\tilde{d}^4 - \tilde{H}^2\tilde{d}^3 + 2\left[\mu(1-\mu)\right]I_S\tilde{d}^2 + I_S & = & 0
\end{eqnarray}
where we recall that $I_S = 0.4 \left[ \mu^{5/3} + (1-\mu)^{5/3}\right]$. A few transformations are made to reduce this equation into a simpler form. First, define the variable $u = \left[\mu(1-\mu)\right]^2 \tilde{d} / \tilde{H}^2$, which changes the polynomial to
\begin{eqnarray}
	u^4 - u^3 + 2\sigma u^2 + \sigma^2 & = & 0
\end{eqnarray}
where $\sigma = \frac{I_S \left[\mu(1-\mu)\right]^3}{\tilde{H}^4}$. Assuming that $u$ is positive and real, this equation can be simplified to 
\begin{eqnarray}
	u^2 + \sigma & = & u^{3/2}
\end{eqnarray}
Then, introducing the variable $X^2 = u$ gives a simplification to the system
\begin{eqnarray}
	X^4 - X^3 + \sigma & = & 0
\end{eqnarray}
followed by $X = 1 + \sigma Y$, which yields the final form
\begin{eqnarray}
	\sigma^3 Y^4 + 3 \sigma^2 Y^3 + 3\sigma Y^2 + Y + 1 & = & 0
\end{eqnarray}
This form is convenient for expanding the solution in terms of the parameter $\sigma$. We note that this expansion is formal and do not try to prove that it converges, rather it will provide us an asymptotic solution. The test solution we use is
\begin{eqnarray}
	Y & = & - \left[ 1 + \alpha_1\sigma + \alpha_2 \sigma^2 + \ldots \right] 
\end{eqnarray}
This is convenient as the higher orders of the solution do not need to be expanded to as high of an order, due to the multiplication by higher powers of $\sigma$. Carrying out the expansion yields the following terms
\begin{eqnarray}
	\alpha_1 & = & 3 \\
	\alpha_2 & = & 15 \\
	\alpha_3 & = & 91 \\
	\alpha_4 & = & 612 \\
	\alpha_5 & = & 4389
\end{eqnarray}

Undoing the transformation then yields a formula for the equilibrium distance, 
\begin{eqnarray}
	 \tilde{d} & = & \frac{\tilde{H}^2 }{ \left[\mu(1-\mu)\right]^2} \left(1-\sigma \left[ 1+ \sum_{i=1}^N \alpha_i \sigma^i\right]\right)^2
\end{eqnarray}
where
\begin{eqnarray}
	\sigma & = & \frac{I_S \left[\mu(1-\mu)\right]^3}{\tilde{H}^4}
\end{eqnarray}

\clearpage
 
\subsection*{Asteroid Data}

The asteroid data given in Tables 1 and 2 were computed using data shown in Tables 3 and 4. 
Data in Table \ref{tab:ast_data} was taken from a variety of sources, including some that were amalgams of observational results. Relevant citations are given in the table. 
Data in Table \ref{tab:pair_data} was taken from Tables 1 and 2 in \cite{pravec2019asteroid}. The mass fractions were computed following the formula given in that paper: given the difference between absolute magnitudes of the two bodies, denoted as $\Delta H$, the mass ratio is computed as $q = 10^{-0.6 \Delta H}$ and the mass fraction as $\mu = q / (1+q)$. Only S-Type asteroid pairs with spin data on both components were considered, yielding the few specific pairs looked at. This allows for a uniform density choice and additional information for the total system energy.

\begin{table}[h!]
\caption{Data for single, contact and binary asteroids. Starred density values are assumed based on spectral type, others have been determined from mission data or from estimates of the total mass and size of an orbiting system. }
\begin{center}
\begin{tabular}{ | c | c || c | c | c | c | c | c |}
\hline
  Asteroid	&  		& Mass  	&  Density		& Primary	& Secondary	& Orbit   & \\
  Name	& Type 	&  Ratio 	& 	 		&  Period  	&  Period 		&  Period  	& Citation \\
		&  		&  $\mu$ 	&  g/cm$^3$ 	&  hr 	&  hr 		&  hr 	& 	\\
\hline
\hline
Bennu & Single & 1 & 1.2 & 4.3 & -- & -- & \cite{bennu_dante}	\\
\hline
Ryugu & Single & 1 & 1.2 & 7.6 & -- & -- & \cite{ryugu_watanabe}	\\
\hline 
\hline
Toutatis & CB & 0.176 & 2.5$^*$ 	& 176 & -- & -- & \cite{meyer_ApJL}	\\
\hline 
Itokawa & CB & 0.212 & 1.9	&	12.132 & -- & -- & \cite{fujiwara_science}	\\
\hline 
1996 HW1 & CB & 0.336 & 1.73$^*$	&	8.7624 & -- & -- & \cite{meyer_ApJL}	\\
\hline
Castalia & CB & 0.416 &2.5$^*$ / 2.67$^*$ & 4.095 & -- & -- & \cite{meyer_ApJL}	\\
\hline 
Arrokoth & CB & 0.4216 & 0.235$^*$	&	15.94 & -- & --& \cite{meyer_ApJL}	 \\
\hline
\hline
1996 FG3 & OB & 0.029 & 1.40 & 3.59	& 16.15		& 16.15	& \cite{scheirich_FG3}	\\
\hline
Didymos & OB & 0.011 & 2.79 & 2.26	& 11.9		& 11.9	& \cite{naidu2020radar}	\\
\hline
Moshup & OB & 0.037 & 2.00 & 2.765	& 16		& 16	& \cite{KW4_ostro}	\\
\hline
2000 DP107 & OB & 0.064 & 1.70 & 2.77	& 42		& 42	& \cite{naidu2015radar}	\\
\hline
1991 VH & OB &  0.079 &  1.70 & 2.62	& 13		& 32.57	& \cite{meyer2024earth}	\\
\hline
Hermes & OB & 0.412 & 1.6 & 13.9	& 13.9		& 13.9	& \cite{margot_hermes}	\\
\hline
\end{tabular}

\end{center}
\label{tab:ast_data} 
\end{table}

\begin{table}[h!]
\caption{Data for asteroid pairs.}
\begin{center}
\begin{tabular}{ | c | c || c | c|| }
\hline
Asteroid  & Mass  & Assumed & Period  \\
 Pair & Fraction & Density &  \\
 Names & $\mu$ & g/cm$^3$ & hr  \\
\hline
\hline
Moore-Sitterly		& 0.04448734 & 2.5  & 3.344727  \\
1999 RP27 & & 2.5 & 4.907058  \\
\hline
Rheinland		& 0.10776534 & 2.5 & 4.2737137 \\
Kurpfalz & & 2.5 & 5.877186 \\
\hline
1999 TU95		& 0.07203316 & 2.5 & 3.40815 \\
2001 DO37 & & 2.5 & 7.9 \\
\hline
2001 QH293		& 0.12628958 & 2.5 &  7.1730 \\
2000 EE85 & & 2.5 & 5.400 \\
\hline
\end{tabular}

\end{center}
\label{tab:pair_data}
\end{table}



\clearpage

\bibliographystyle{plain}

\bibliography{../../../bibliographies/biblio_books,../../../bibliographies/biblio_article}

\end{document}